\documentclass[aps,prb,10pt,notitlepage,superscriptaddress,twocolumn]{revtex4-1}

\usepackage{graphicx}
\usepackage{amsmath}
\usepackage{amssymb}
\usepackage{amsfonts}
\usepackage{color}

\newcommand{\asf}{(TMTTF)$_2$AsF$_6$}
\newcommand{\pf}{(TMTTF)$_2$PF$_6$}
\newcommand{\bff}{(TMTTF)$_2$BF$_4$}
\newcommand{\sbf}{(TMTTF)$_2$SbF$_6$}
\newcommand{\cl}{(TMTTF)$_2$ClO$_4$}
\newcommand{\br}{(TMTTF)$_2$Br}

\begin{document}

\title{Electronic properties of Fabre charge-transfer salts under
various temperature and pressure conditions}

\author{A. C. Jacko}\affiliation{Institut f\"ur Theoretische Physik, Goethe-Universit\"at Frankfurt, Max-von-Laue-Stra{\ss}e 1, 60438 Frankfurt am Main, Germany}
\author{H. Feldner} \affiliation{Institut f\"ur Theoretische Physik, Goethe-Universit\"at Frankfurt, Max-von-Laue-Stra{\ss}e 1, 60438 Frankfurt am Main, Germany}

\author{E. Rose}  \affiliation{1. Physikalisches Institut, Universit\"at Stuttgart, Pfaffenwaldring 57, 70550 Stuttgart, Germany}

\author{F. Lissner} \affiliation{Institut f\"ur Anorganische Chemie, Universit\"at Stuttgart, Pfaffenwaldring 55, 70550 Stuttgart, Germany}

\author{M. Dressel}  \affiliation{1. Physikalisches Institut, Universit\"at Stuttgart, Pfaffenwaldring 57, 70550 Stuttgart, Germany}

\author{Roser Valent\'\i} \affiliation{Institut f\"ur Theoretische Physik, Goethe-Universit\"at Frankfurt, Max-von-Laue-Stra{\ss}e 1, 60438 Frankfurt am Main, Germany}
\author{Harald O. Jeschke} \affiliation{Institut f\"ur Theoretische Physik, Goethe-Universit\"at Frankfurt, Max-von-Laue-Stra{\ss}e 1, 60438 Frankfurt am Main, Germany} 

\begin{abstract}
  Using density functional theory, we determine parameters of
  tight-binding Hamiltonians for a variety of Fabre charge transfer
  salts, focusing in particular on the effects of temperature and
  pressure.
  Besides relying on previously published crystal structures, we
  experimentally determine two new sets of structures; {\sbf} at
  different temperatures and {\pf} at various pressures.
  We find that a few trends in the electronic behavior can be
  connected to the complex phase diagram shown by these materials.
  Decreasing temperature and increasing pressure cause the systems to become
  more two-dimensional.  We analyze the importance of correlations by
  considering an extended Hubbard model parameterized using Wannier
  orbital overlaps and show that while charge order is strongly
  activated by the inter-site Coulomb interaction, the magnetic order
  is only weakly enhanced. Both orders are suppressed when the
  effective pressure is increased.
\end{abstract}

\maketitle

\section{Introduction}\label{secI}

\begin{figure}
\begin{center}
 \includegraphics[width=0.95\columnwidth]{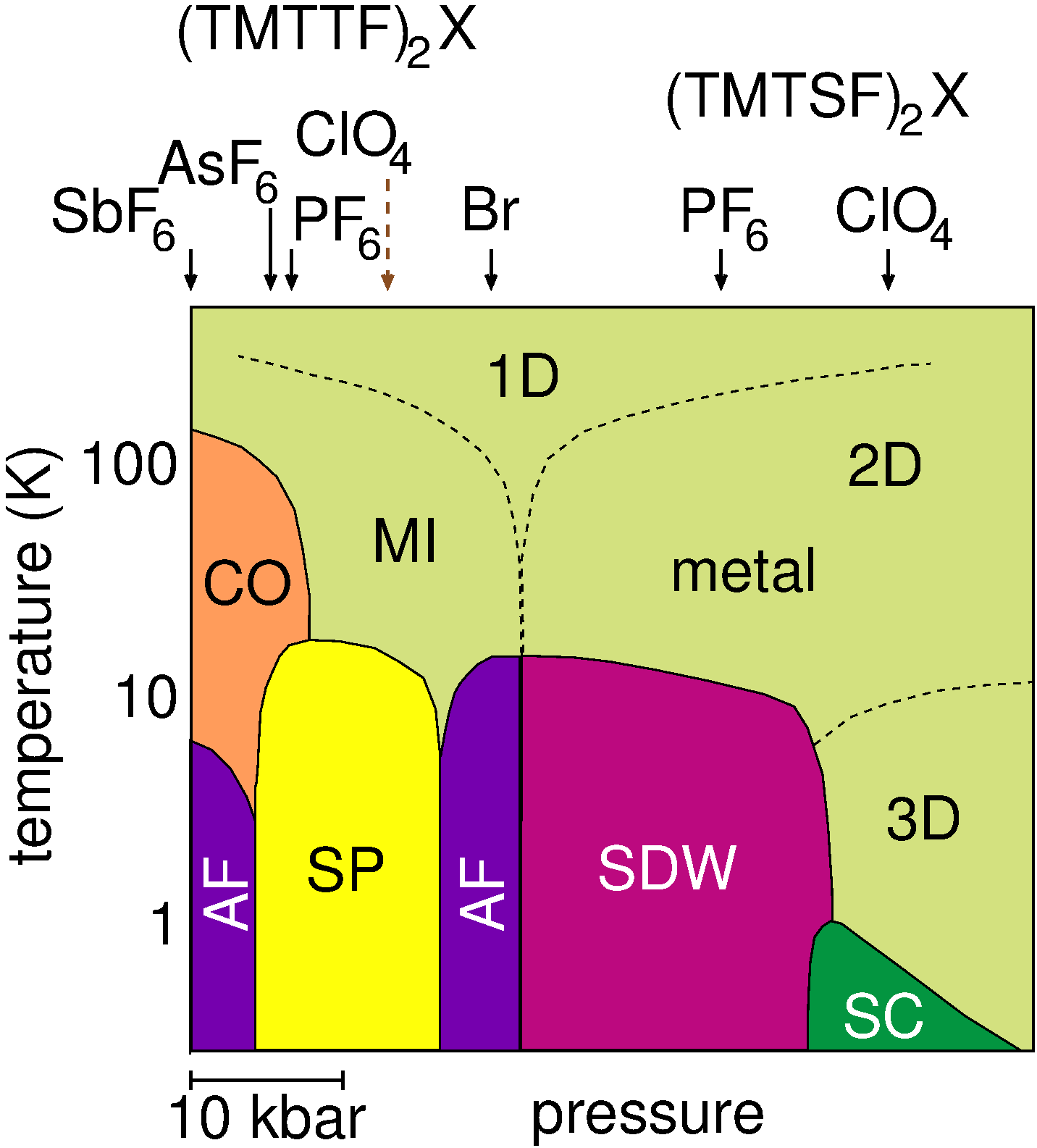}
 \caption{Temperature-pressure phase diagram for the TMTTF and TMTSF
   charge transfer salts, as first suggested in
   Ref. \onlinecite{jerome91}, and refined by many others.  Position
   in the phase diagram can be tuned by physical pressure, or chemical
   pressure (changing anion). The ambient pressure position for each
   salt is indicated with an arrow above the diagram. An increase of
   pressure (external or chemical), causes the system to be less one
   dimensional.  The position of {\cl} is not well known. In the phase
   diagram its approximate position has been indicated with a dashed
   arrow. The possible phases are charge ordered (CO), Mott insulating
   (MI), antiferromagnetic (AF), spin Peierls (SP), spin density wave
   (SDW), superconducting (SC) and 1D, 2D or 3D
   metal.}\label{fig:phasediagram}
\end{center}
\end{figure}

Quasi-one dimensional organic salts formed from
tetramethyltetrathiafulvalene (TMTTF) molecules - also known as Fabre
charge-transfer (CT) salts - have been intensively investigated in the
last two decades since they exhibit a rich variety of phases like 
 antiferromagnetism, superconductivity, charge ordering,
spin-density wave ordering or spin-Peierls
behavior.~\cite{jerome91,mori06,yasuzuka09} Such phases can be driven
both by external (physical) pressure as well as by chemical pressure
(see Fig.~\ref{fig:phasediagram}).  Even though a few successful models
have been proposed for the description of these systems, constructing
a consistent microscopic picture of the relationships between the
various phases remains a
challenge.\cite{kawakami03,yu04,doironleyraud10}

The primary avenue of the present work is to understand the
microscopic origin of the close competition between the different
phases in these compounds as a function of chemical and external
pressure as well as temperature. For that, we performed \textit{ab
  initio} density functional theory (DFT)~\cite{kandpal09,jeschke12}
 and model Hamiltonian
calculations for several Fabre CT salts whose crystal structures were
 determined at different temperatures and pressure
and investigated variations of
their electronic properties with temperature and pressure.\footnote{}
By computing the real-space overlaps of Wannier orbitals for the bands
near the Fermi level, we parametrize a two-band tight-binding
Hamiltonian model for the various systems and examine the differences
in their electronic hopping parameters.  In this way, we can connect
structural and chemical modifications with changes in the electronic
properties.  Furthermore, in order to analyze some of the preferred
orderings we consider a description of the Fabre CT salts in terms of
an extended Hubbard Hamiltonian including on-site and inter-site
Coulomb interaction terms. The kinetic part of this model is given by
the computed hopping parameters for the various compounds.  We discuss
spin-spin and charge-charge correlation properties by diagonalizing
the model.

The work is organized as follows. Sections~\ref{secII} and
\ref{secIII} are dedicated to the description of the computational
details as well as the crystal structure of the Fabre CT salts.
 In Sections~\ref{secIV}, \ref{secV} and \ref{secVI} we
present, respectively, our results on the electronic structure,
tight-binding models as well as extended Hubbard models for a few members
of Fabre CT salts. Discussion and conclusions are given in
Sections~\ref{secVII} and \ref{secVIII}.

\section{Computational details}\label{secII}

The electronic structure calculations presented here were performed in
an all-electron full-potential local orbital basis using the FPLO
package.\cite{koepernik99} The densities were converged on an $(8
\times 8\times 8)$ $k$ mesh, using a generalized gradient
approximation (GGA) functional.\cite{perdew96}

For materials without published hydrogen coordinates, hydrogen atoms
were placed according to the expected bond lengths and angles of a
methyl group (C--H distance 1.1 \AA, C--C--H angle 109$^\circ$).  With
the bond length and angle fixed, one has the freedom to choose the
rotation angle of the set of hydrogen atoms on each methyl group
around the C--C bond. We chose this angle such that one hydrogen is as
far out of the plane of the molecule as possible.  We tested the
effect of this choice on the band structure and found no contribution
to the bands of interest near the Fermi level since in this energy
region only TMTTF bands of $\pi$ origin are involved.

Some of the structures had suspicious bond lengths and angles, 
therefore the atomic coordinates were relaxed with DFT using the Vienna
\textit{ab initio} Simulation Package (VASP, version
5.2.11),~\cite{kresse93,kresse96} with a projector-augmented wave
basis.~\cite{blochl94,kresse99} We used the GGA functional,\cite{perdew96} and included Van
der Waals corrections~\cite{grimme06} for the relaxations.  We
performed two kinds of relaxations; in one relaxation we kept sulfur
and the heavy anion atoms coordinates fixed, and in the other relaxation all
atom positions were relaxed. The differences between the two relaxed
structures were minimal.
The atomic coordinates were converged to an energy difference of 1 meV
on a $(5 \times 5 \times 5)$ $k$-mesh, with a plane wave cutoff energy
of 500 eV.

The tight-binding parameters were obtained by constructing Wannier
orbitals for the TMTTF bands at the Fermi level and computing
real-space overlaps, as implemented in FPLO.  Another way to generate
these parameters is to fit the band structure of the model Hamiltonian
to the DFT bands. The latter method can become difficult when many
hopping parameters need to be fitted; there can be a number of
solutions which reproduce the DFT bands equally well, but differ in
physical details (such as relative strengths of certain bonds).  By
using Wannier orbital overlaps we can be sure that our parameter
values have a clear physical interpretation.

The exact diagonalization of the extended Hubbard model was performed
by considering system sizes of $4\times4$ TMTTF sites with periodic boundary
conditions.

\section{Crystal Structure}\label{secIII}

\begin{figure}
\begin{center}
\includegraphics[width=0.95\columnwidth]{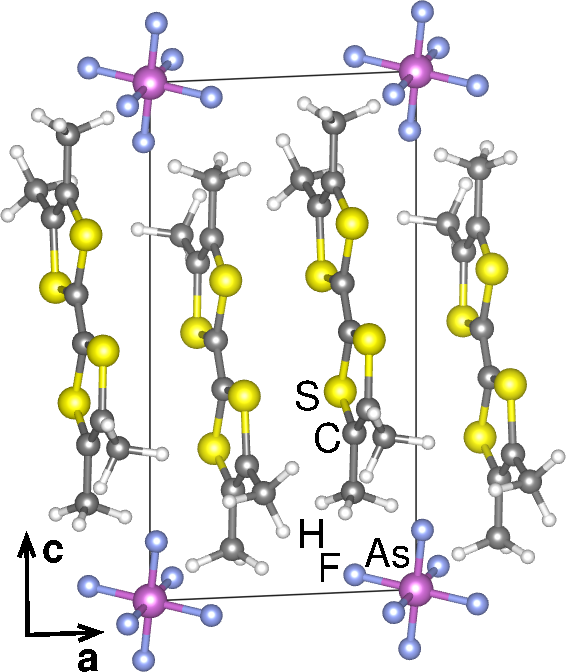}
 \caption{Crystal structure of the Fabre CT salts projected into the
   $ac$ plane.  The organic molecules form $\pi$-stacked 1D chains
   along the crystal $a$ direction (with a slight zig-zag pattern),
   and form layers parallel to the $ab$ plane.  These organic layers alternate
   with anion layers (with the anions centered on the pink As sites)
   stacked in the $c$ direction. Grey atoms are carbon, yellow are
   sulfur, while hydrogen atoms are shown in white.}\label{fig:tmttfcrystal}
\end{center}
\end{figure}

The Fabre CT salts consist of alternate layers of TMTTF molecules
(cations) and monovalent anions, stacked in the $c$ direction (see
Fig.~\ref{fig:tmttfcrystal}).  In between the cation layers, the
planar TMTTF molecules form $\pi$-stacked one dimensional chains in
the $a$ direction with a slight `zig-zag' arrangement. There is a
charge transfer of one electron from each (TMTTF)$_2$ dimer to each
anion, {\it i.e.} the TMTTF molecules carry half a
hole on average. There are two classes of anions: those that conform to the
$P\,\bar{1}$ symmetry of the TMTTF part of the crystal (such as PF$_6$), and those that
break that symmetry (such as ClO$_4$).  Also, the anion species influence the
proportions of the unit cell, as well as the intra- and inter-chain
hopping strengths.  The inter-chain hopping strengths are not only
determined by the distance between the TMTTF molecules, but also by
changes in their zig-zag arrangement 
(that is to say, how far away from a perfectly aligned stack they are, and in what direction).

TMTTF molecules within a chain show a
slight dimerization along the chain.  We can quantify this structural
dimerization as the difference between the larger dimerization
distance of adjacent TMTTF molecules, $d_1$, and the shorter
dimerization distance, $d_0$, normalized by the sum of the two
distances:
\begin{equation} \label{eq:dimstruc}
 \partial_{struc} = 2\frac{d_1-d_0}{d_1+d_0},
\end{equation}
These distances are defined as the distances between the centres of mass of the C and S atoms in each TMTTF molecule.

Table~\ref{tab:properties} shows the
structural dimerization of the materials investigated in this
work. This table also includes the electronic dimerization, which is
introduced in Sec. V.
In general, the structural dimerization increases  slightly
with increasing temperature.

Cooling from room temperature to T = 4 K, {\asf} and {\pf} both show charge ordering
phase transitions and spin-Peierls transitions. 
These ordering transitions are not visible in the crystal structures; there are no significant changes in the structural dimerization from T= 300 K to T= 4 K. {\sbf} shows similar changes in the structural dimerization between 100 K and room temperature, and does not go through any ordering transitions in this range.

\begin{table}
\begin{tabular}{|l|ccc|}
\hline
\textbf{Anion} & $\partial_{struc}$ & $\partial_{elec}$ & Ref. \\ \hline
 SbF$_6$ (100\,K) 		& 0.007 	& 0.042 & new \\
 SbF$_6$ (140\,K, sample 1) 	& 0.011 	& 0.067 & new \\
 SbF$_6$ (140\,K, sample 2) 	& 0.013 	& 0.094 & new \\
 SbF$_6$ (180\,K)		& 0.020 	& 0.115 & new \\
 SbF$_6$ (200\,K) 		& 0.023 	& 0.141 & new \\
 SbF$_6$ (300\,K, sample 1)  	& 0.047 	& 0.279 &  new    \\ 
 SbF$_6$ (300\,K, sample 2) 	& 0.041 	& 0.298 & new \\ \hline
 AsF$_6$ (4\,K)  		& 0.007 	& 0.100 & \onlinecite{granier88}  \\ 
 PF$_6$ (4\,K)  		& 0.009 	& 0.126 & \onlinecite{granier88}   \\ 
 AsF$_6$ (300\,K)  		& 0.041 	& 0.110 &  \onlinecite{liautard82b}*  \\ 
 PF$_6$  (300\,K) 		& 0.040 	& 0.230 &  new \\
 PF$_6$ (300\,K, 0.3\,GPa)  	& 0.018 	& 0.577 &  new        \\ 
 PF$_6$ (300\,K, 0.6\,GPa)  	& 0.016 	& 0.595 &   new       \\ 
 PF$_6$ (300\,K, 0.9\,GPa) 	& 0.002  	& 0.477 &   new       \\ 
 PF$_6$ (300\,K, 1.5\,GPa) 	& 0.003 	& -0.454 &   new      \\ 
 PF$_6$ (300\,K, 2.0\,GPa) 	& 0.010 	& -0.397 &   new     \\ 
 PF$_6$ (300\,K, 2.7\,GPa) 	& 0.024 	& -0.183 &   new     \\ \hline
 Br  (300\,K) 			& 0.019 	& -0.189 &  \onlinecite{galigne78}*  \\ 
 ClO$_4$ (300\,K)  		& 0.040 	&  0.616 &    \onlinecite{liautard84}*    \\ 
 BF$_4$ (100\,K)  		& 0.020 	& -0.054 &  \onlinecite{galigne79a}  \\ 
 BF$_4$ (300\,K)  		& 0.028 	&  0.336 &     \onlinecite{galigne79a}  \\  \hline
\end{tabular} 
\caption{Structural and electronic dimerization of the Fabre CT salts
considered in the present work.  The structural dimerization of the
TMTTF molecules is defined in Eq.~\eqref{eq:dimstruc}, and the
electronic dimerization is defined in Eq.~\eqref{eq:dimelec} in
Sec. V. References marked with * have no (or
unrealistic) published hydrogen coordinates.  Note that ClO$_4$ and
BF$_4$ are tetrahedral anions, and so do not conform to the reported
$P\,\bar{1}$ symmetry (the anions do not have the required inversion
symmetry).  The anion ordering transition only occurs in these systems
where the anion does not have inversion symmetry.  The change in sign
of $\partial_{elec}$ indicates that the shorter bond has the smaller
$t$.
}\label{tab:properties}
\end{table}

\section{Electronic Structure}\label{secIV}

\begin{figure}
\begin{center}
\includegraphics*[width=0.47\textwidth]{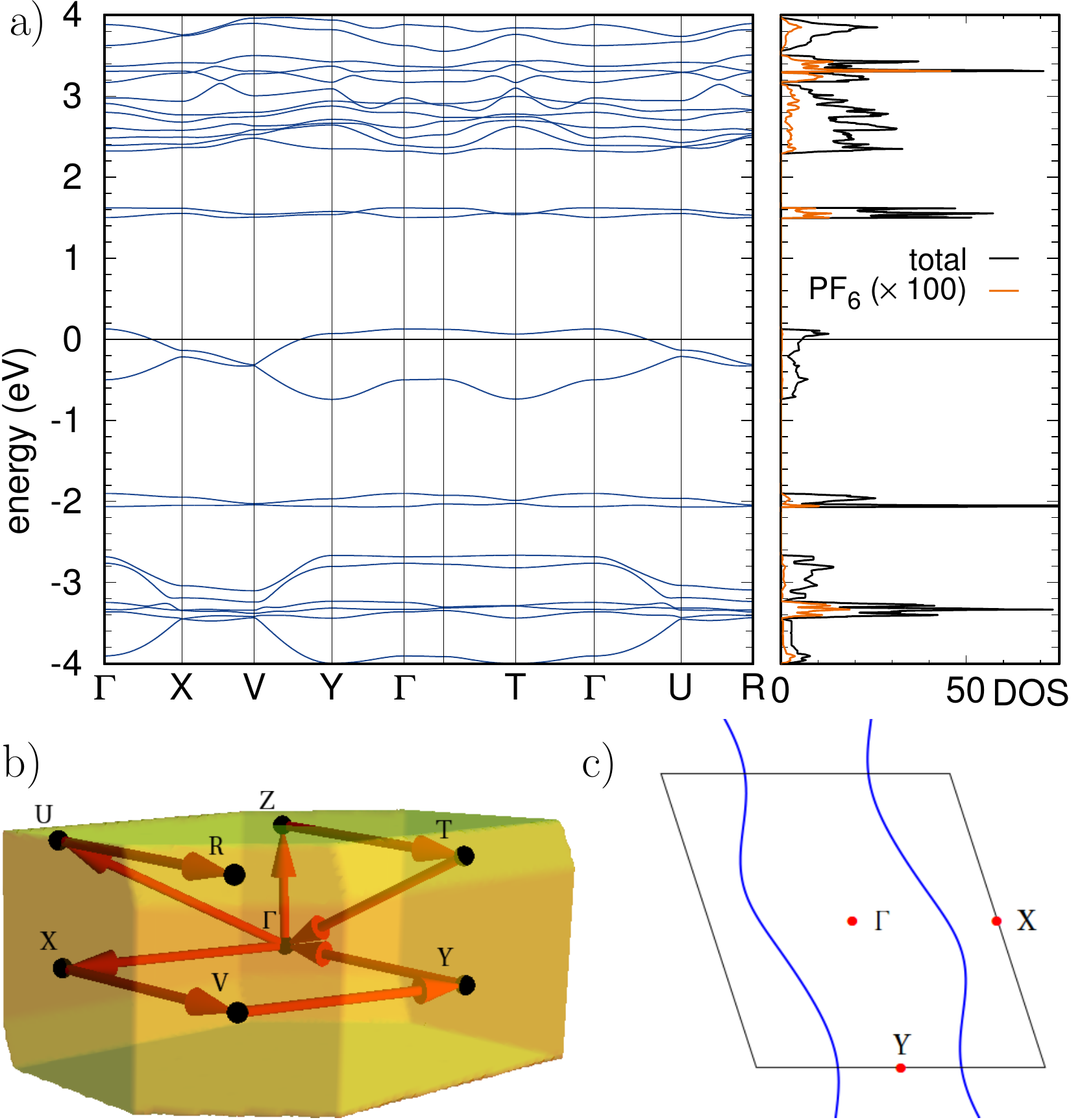}
\caption{Electronic properties 
of {\pf}  ($T=4$~K structure). a) Band structure and density of states.
b) Path through $k$-space considered for the band structure plotting.
  c) Fermi surface in the $k_z = 0$
  plane .  The total density of states is shown in black and the
  partial density of states of the anions (increased by a factor of
  100) is shown in red (dashed). The partial density of states shows
  that within this energy window, all of the bands in this energy
  window have predominantly TMTTF character, and the two bands at the
  Fermi level are nearly purely TMTTF.}\label{fig:pf6bands}
\end{center}
\end{figure}

In the following we examine the electronic properties of {\pf} in
detail and will use this analysis as a baseline for understanding the
Fabre CT salts. In Fig.~\ref{fig:pf6bands} (a) we present the band
structure and density of states of {\pf} in a window of energy
$[-4\,{\rm eV},4\,{\rm eV}]$ around the Fermi level. The bands have
been drawn along the high symmetry path shown in the Brillouin zone in
Fig.~\ref{fig:pf6bands} (b).  It is clear from the partial density of
states that near the Fermi level all the bands are predominantly due
to the TMTTF molecules.  In fact, the nearest anion bands are about
4.1 eV below the Fermi level, and more than 10 eV above it.  The two
$3/4$-filled organic bands near the Fermi energy are a common feature
of the Fabre salts as a result of hole doped pairs of TMTTF molecules.
In {\pf} these bands are well separated from the rest of the bands,
with gaps of more than 1 eV to the lower valence bands and upper
conduction bands respectively.  The size of the gaps vary with anion
type, and sometimes anion bands cross the two TMTTF bands (as in the case of {\br}).

The quasi-one dimensionality of this system is manifested in the
band structure (Fig.~\ref{fig:pf6bands} (a)) where we find very little
dispersion in the $k_y$ and $k_z$ directions and bands only cross the
Fermi level in the $k_x$ direction.  This can be also observed in the
Fermi surface cut at $k_z=0$ shown in Fig.~\ref{fig:pf6bands} (c).
This quasi-one dimensional behavior is a typical feature of the Fabre
CT salts.

\begin{figure}
\begin{center}
 \includegraphics[width=0.95\columnwidth]{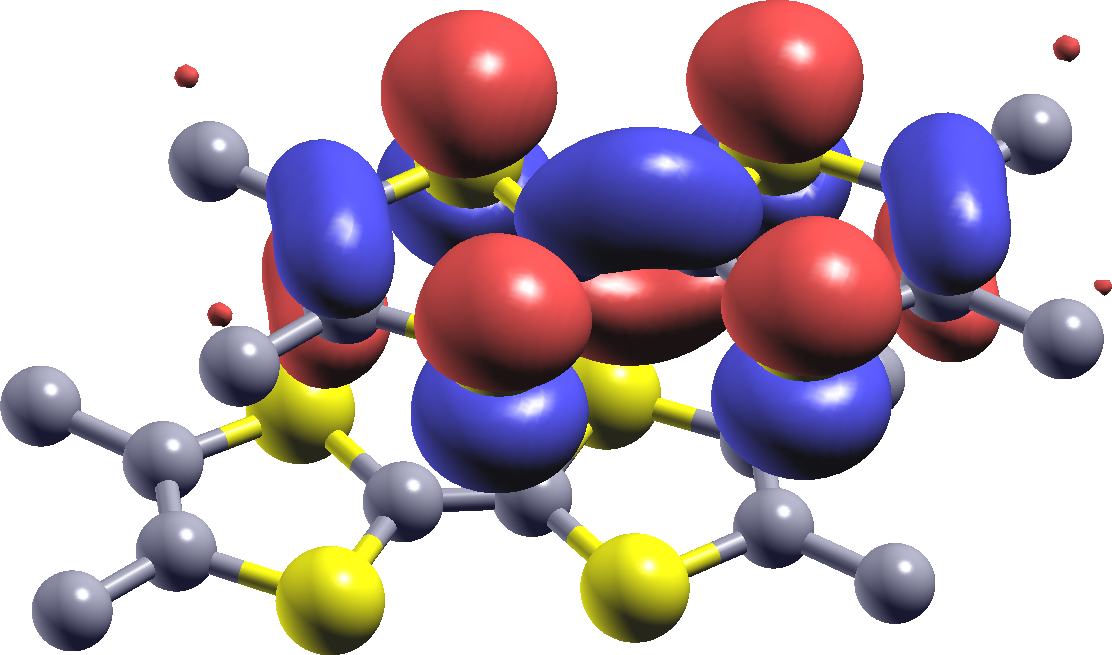}
 \caption{Wannier orbital for the {\pf} bands near the Fermi level.
   It is clear that this Wannier orbital has the symmetry of the HOMO
   of a TMTTF molecule in the gas phase.  The other Wannier orbital
   needed to describe the two organic bands corresponds to the HOMO of
   the second TMTTF molecule in the unit cell (and is related to the
   first one by inversion symmetry).  This is in agreement with the
   information in the partial density of states
   (Figure~\ref{fig:pf6bands} (a)); the orbitals near the Fermi energy
   are predominantly of TMTTF nature.} \label{fig:pf6wf}
\end{center}
\end{figure}

In order to further characterize the electronic structure of these
systems, we generate Wannier orbitals for the two organic bands near
the Fermi level as described in Section~\ref{secI}. An example is
shown in Fig.~\ref{fig:pf6wf}. These bands have the symmetry of the
TMTTF highest occupied molecular orbitals (HOMOs), partially
depopulated by the charge transfer of one electron from a pair of
TMTTF molecules to the anion layer. These two bands determine the
low-energy physics of these systems and, in what follows, we shall
concentrate on the analysis of this band manifold.
We note that we are not considering DFT calculations beyond GGA and
therefore leave correlation effects (beyond GGA) to be explicitely treated in the
model calculations.

\section{Tight-Binding Model}\label{secV}
Wannier orbitals form a natural basis for a tight-binding model. By 
computing overlaps between the orbitals, we can 
parameterize the two HOMO bands at the Fermi energy in terms of a two-site
tight-binding Hamiltonian where the lattice sites are defined as the
centers of mass of the two TMTTF molecules in each unit cell:
\begin{equation}
 \hat{H}_N = \mu\sum_i c_{i}^\dagger c_{i}-\sum_{ \langle  i,j \rangle_N} t_{ij} c_{i}^\dagger c_{j}, \label{eq:modelH}
\end{equation}
$\mu$ is the on-site energy, $t_{ij}$ are hopping parameters
between sites $i$ and $j$ and the sum over $\langle i,j \rangle_N$
indicates that only hoppings up to the $N^{th}$ nearest neighbor are
included. In listing hopping parameters, we will use 
\begin{equation}
 t_{ij}\equiv t_{\alpha}(r_{ij})
\label{t_hop}
\end{equation}
where $r_{ij}$ are distances between TMTTF centers of mass and
$\alpha=0,1,2,\dots$ counts neighbour distances in ascending order.
In the discussion that follows, we include hoppings up to the 8$^{th}$
nearest neighbor ($N=8$). These 8 hopping terms do not include any
inter-layer hopping, and therefore the resulting tight-binding bands
have no dispersion in the $k_z$ direction.  The resulting
tight-binding parameters for the eight shortest inter-site distances
are shown in Table~\ref{tab:hoppings}. The longer hopping terms are of the same order as $t_7$ or smaller.

\subsection{Anion dependence of the structural and electronic
properties}\label{secVa}

\begin{figure}
\begin{center}
 \includegraphics[width=0.95\columnwidth]{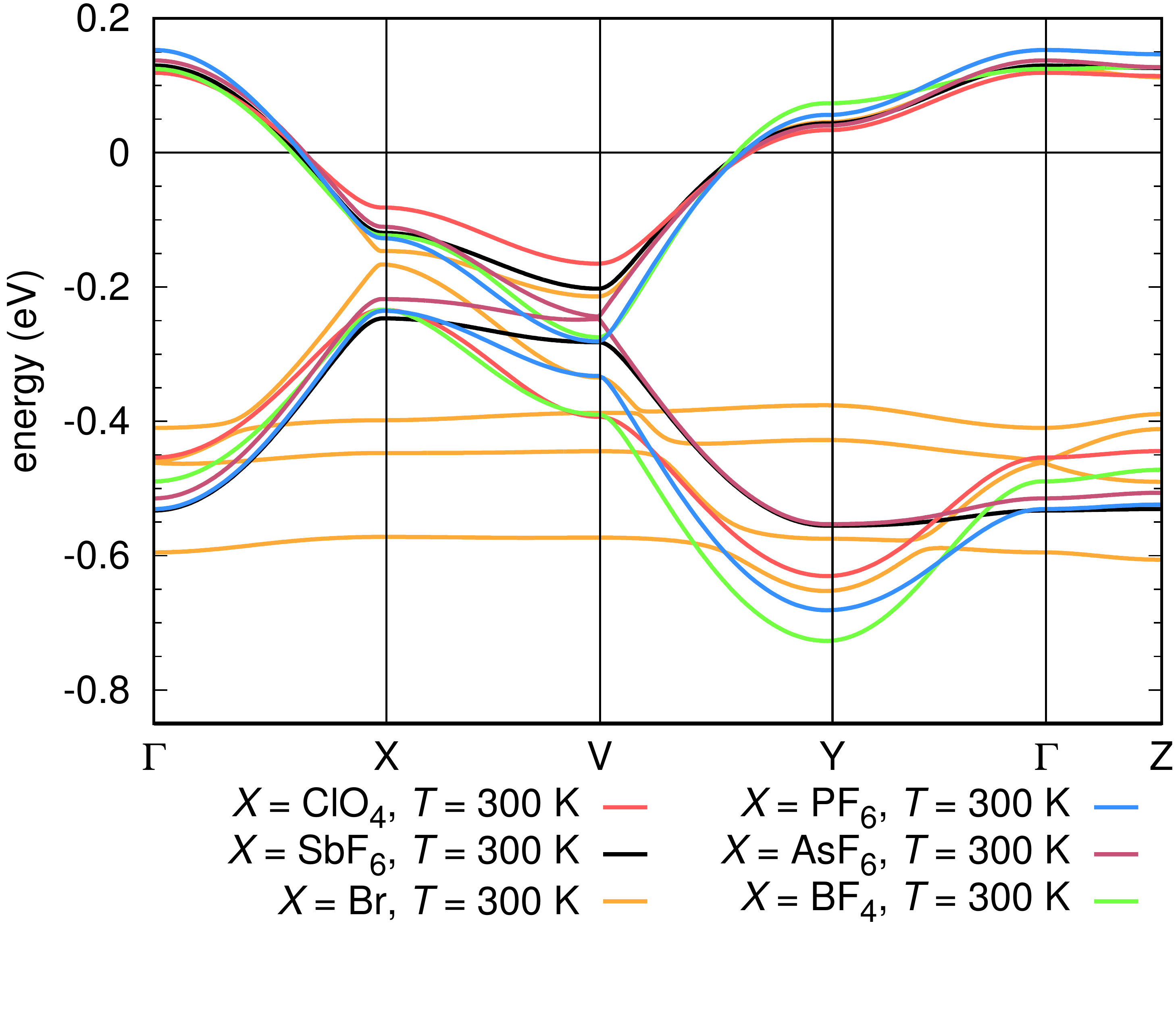}
 \caption{Band structures in the energy window
[-0.8eV,0.2eV] of the TMTTF salts with crystal structures
measured at ambient pressure and
   room temperature. In this energy range all of the materials shown
   here have two bands arising from TMTTF HOMO orbitals. The {\br}
   salt additionally has three anion bands within this window. The
   common TMTTF bands differ in details; between $\Gamma$ and X
   (corresponding to the in-chain direction) the bands are very
   similar. There is more variation in the inter-chain direction,
  indicating the differing degrees of inter-chain
   coupling.}\label{fig:bs_All_RT}
\end{center}
\end{figure}

In Figure~\ref{fig:bs_All_RT} we show
 the band structure  of the
various Fabre CT salts considered in this study with crystal
structures measured at ambient pressure
and room temperature (see Table~\ref{tab:properties}).
  This comparison allows us to analyze
the effects of chemical pressure (i.e. anion substitution) on the
electronic properties.  For {\br} there are three additional Br bands
crossing the lower organic band.

We observe that the TMTTF bands vary only modestly with anion  at a given
temperature, particularly at the $\Gamma$ and $Z$ points.  The largest
difference in the band structure is seen along the $X$-$V$ path,
($X=(0.5, 0, 0)$, $V=(0.5, 0.5, 0)$ in units of the reciprocal lattice
vectors) where the indirect influence of the anion is most prominent.
It is clear that in {\br} strong mixing with the Br bands distorts the
TMTTF bands around the avoided crossings.  Away from the avoided
crossings the bands are similar to the TMTTF bands observed for the
other salts. It is worth noting that {\br} is the only salt studied
here with easily accessible metallic and superconducting
states.\cite{pedron94}

\begin{figure*}
\includegraphics[width=0.8\textwidth]{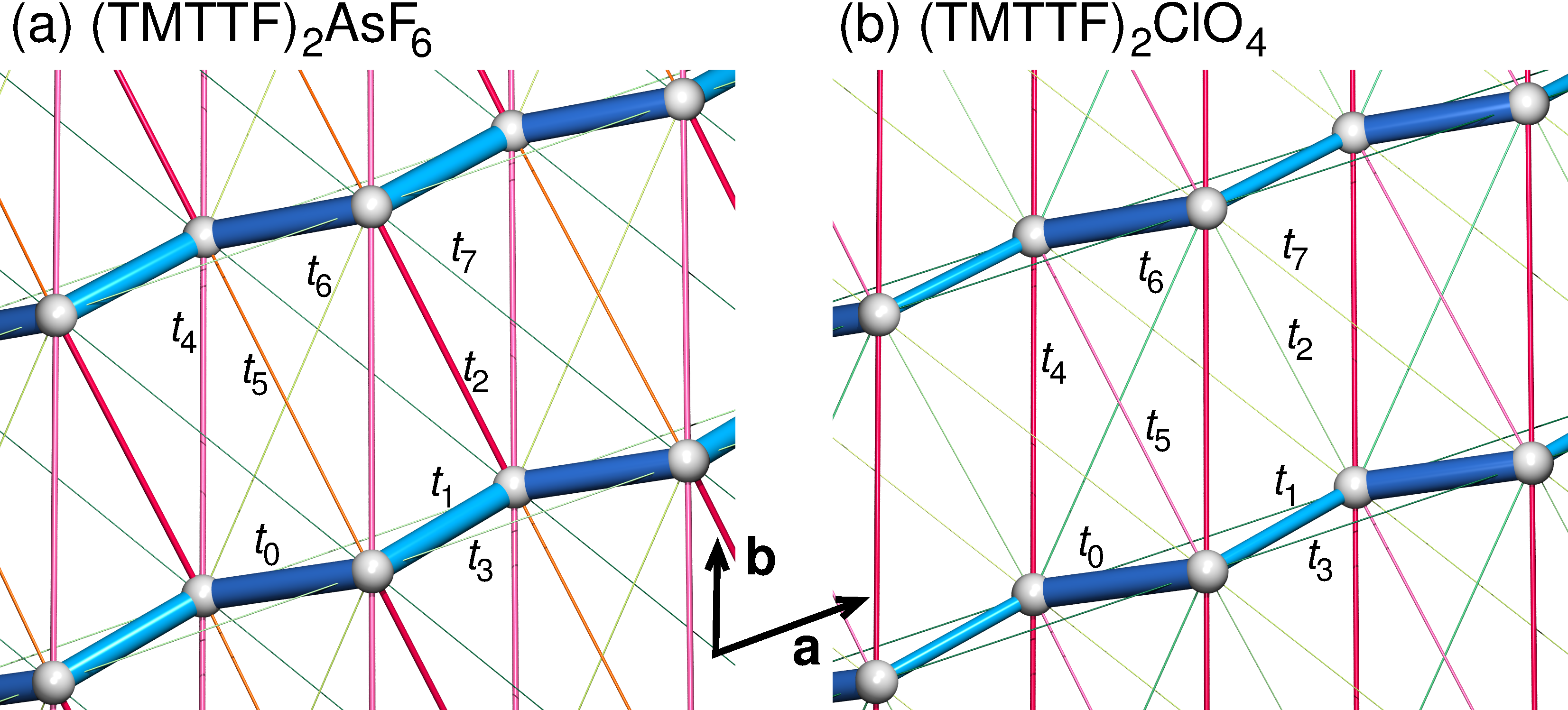}
\caption{Visualization of the strength of the hopping between the
  sites of the tight-binding model, the centers of mass of the TMTTF
  molecules (gray spheres); shown for (a) {\asf} and (b) {\cl} 
  (both at room temperature). The diameter of the bonds is
  proportional to the tight binding parameter strength $|t_\alpha|$.
  $|t_\alpha|$ above 0.1 eV are each a different shade of blue, $|t_\alpha|$
  between 0.1 eV and 0.01 eV are a shade of red/orange, while $|t_\alpha|$
  less than 0.01 eV are a shade of green.  See
  Table~\ref{tab:hoppings} in the Appendix for $t_\alpha$
  values.}\label{fig:TBH_anion}
\end{figure*}

In Fig.~\ref{fig:TBH_anion} we show the real-space network of hopping
terms $t_{ij}$ (see Eq.~\ref{t_hop}) 
between TMTTF molecules computed from the Wannier
orbitals overlaps for {\asf} (Fig.~\ref{fig:TBH_anion} (a)) and {\cl}
(Fig.~\ref{fig:TBH_anion} (b)). The strength of the hopping is linearly encoded
into the bond diameter. This figure shows that these materials have a
preferred hopping direction (the direction with the thickest bonds), forming
one dimensional chains. The ratio of inter-chain to intra-chain
hopping strengths is smaller in {\asf} than in {\cl} which indicates
that {\asf} is more one-dimensional.

While all of the materials have strong intra-chain hopping terms
(of order $\sim$0.15-0.25~eV)), the inter-chain hopping terms can vary by
about an order of magnitude (see Table~\ref{tab:hoppings} in the
Appendix). The values of the intra-chain hopping parameters in our
work are consistent with those found for similar systems in previous
experimental and theoretical
investigations.~\cite{grant82,whangbo82,ducasse86,pedron94,pouget96,dumm00,yoshimi12}
Missing in those previous studies is a thorough analysis of the
intra-chain dimerization as well as the inter-chain hopping
parameters.

In Table~\ref{tab:properties} we quantify the electronic dimerization
for the Fabre CT salts studied in this work analogously as we did for
the structural dimerization, i.e.
\begin{equation} \label{eq:dimelec}
\partial_{elec} = 2\frac{t_0-t_1}{t_0+t_1}
\end{equation}
where $t_0$ ($t_1$) is the hopping term
corresponding to the smallest (second smallest) bond length.
While we observe a significant dependence on the
nature of the anion, the structural and electronic dimerizations seem
to be uncorrelated. This can be understood physically:
the electronic dimerization is defined by hopping integrals whose
magnitude depends on the orientation of the overlapping orbitals as
well as on their separation. If the orientation is more favorable
along the longer intra-chain bond, then the more distant overlap can
be larger. This is the case for the structures of
 {\pf} above $P=0.9$~GPa, {\bff} at
$T=100$~K, and {\br} at room temperature; a negative value of
$\partial_{elec}$ in Table~\ref{tab:properties} indicates that the
longer bond has a larger hopping strength.  Focusing on the anions
with octahedral symmetry at room temperature, we observe that while
the structural dimerization has a consistent trend downwards as the
anion changes from (SbF$_6$)$^-$ through (AsF$_6$)$^-$ to (PF$_6$)$^-$
(chemical pressure, smaller volume) and then further downwards as
pressure is applied, the electronic dimerization follows the opposite
trend. We will discuss this behavior below.

\subsection{Temperature dependence of the structural and electronic
properties}\label{secVb}

We proceed now with the analysis of the temperature dependence of
the structural and electronic behavior of 
a few Fabre CT salts. This study is done
by performing ground state DFT calculations for structures determined at different
temperatures.
The investigation for {\asf} and {\pf} is done by
considering crystal structures obtained experimentally at $T=4$~K and
at $T=300$~K.  The investigation for {\sbf} is done using
crystal structures determined experimentally at temperatures between
$T=100$~K and 300~K.

\begin{figure}
\begin{center}
 \includegraphics[width=0.95\columnwidth]{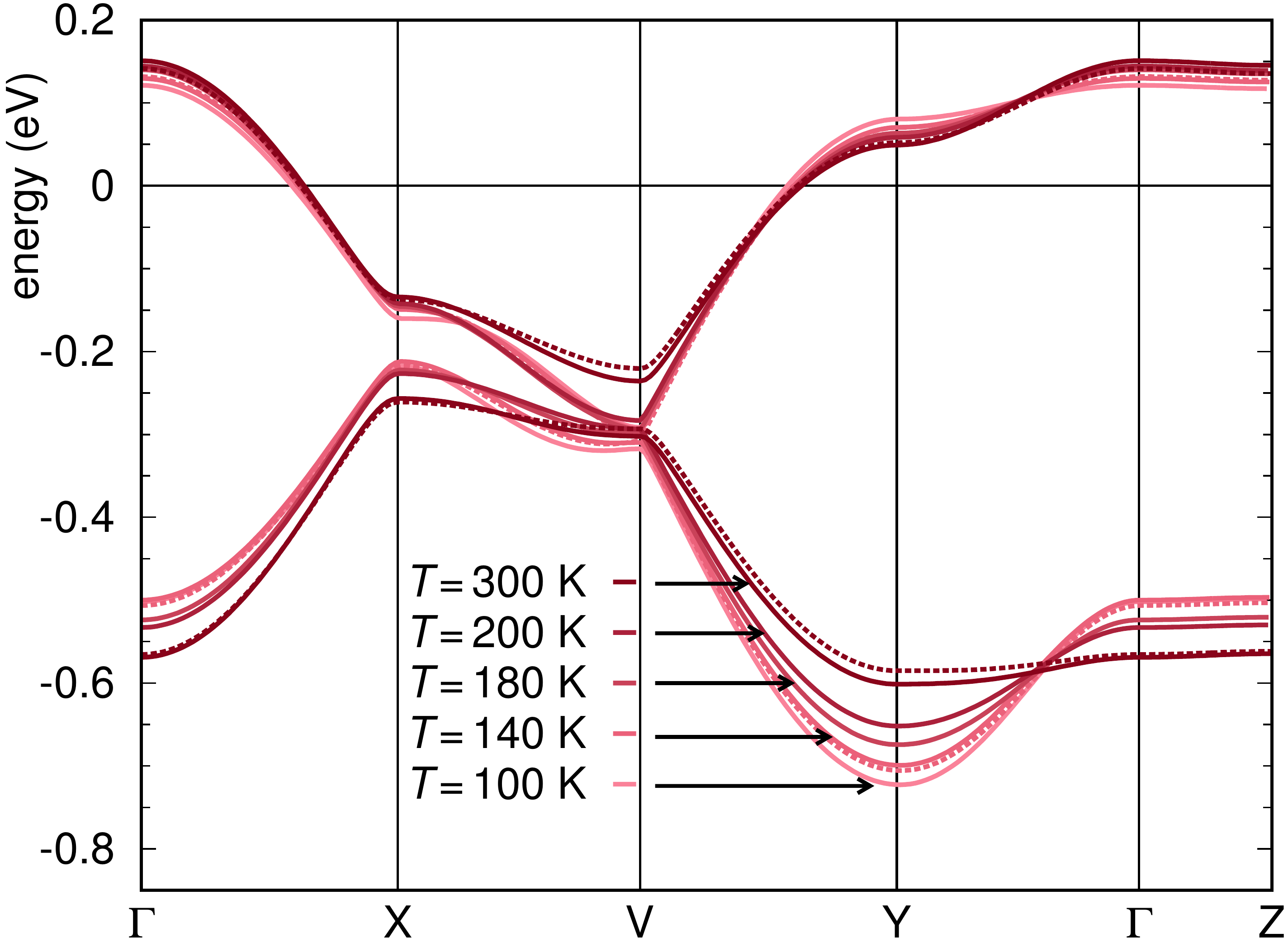}
 \caption{Band structure of {\sbf} calculated from the crystal structures
   obtained at several temperatures between $T=100$~K and 300~K. At
   $T=140$~K and 300 K, structures from two different samples were
   used; the additional bands at those temperatures are plotted with
   dashed lines. As the temperature is decreased, the bandwidth
   increases and the dispersion between the $X$ and $V$ points becomes
   steeper; this indicates that the electronic structure becomes more
   two dimensional with decreasing temperature.}\label{fig:bssbf6temp}
\end{center}
\end{figure}

\begin{figure}
\begin{center}
 \includegraphics[width=0.95\columnwidth]{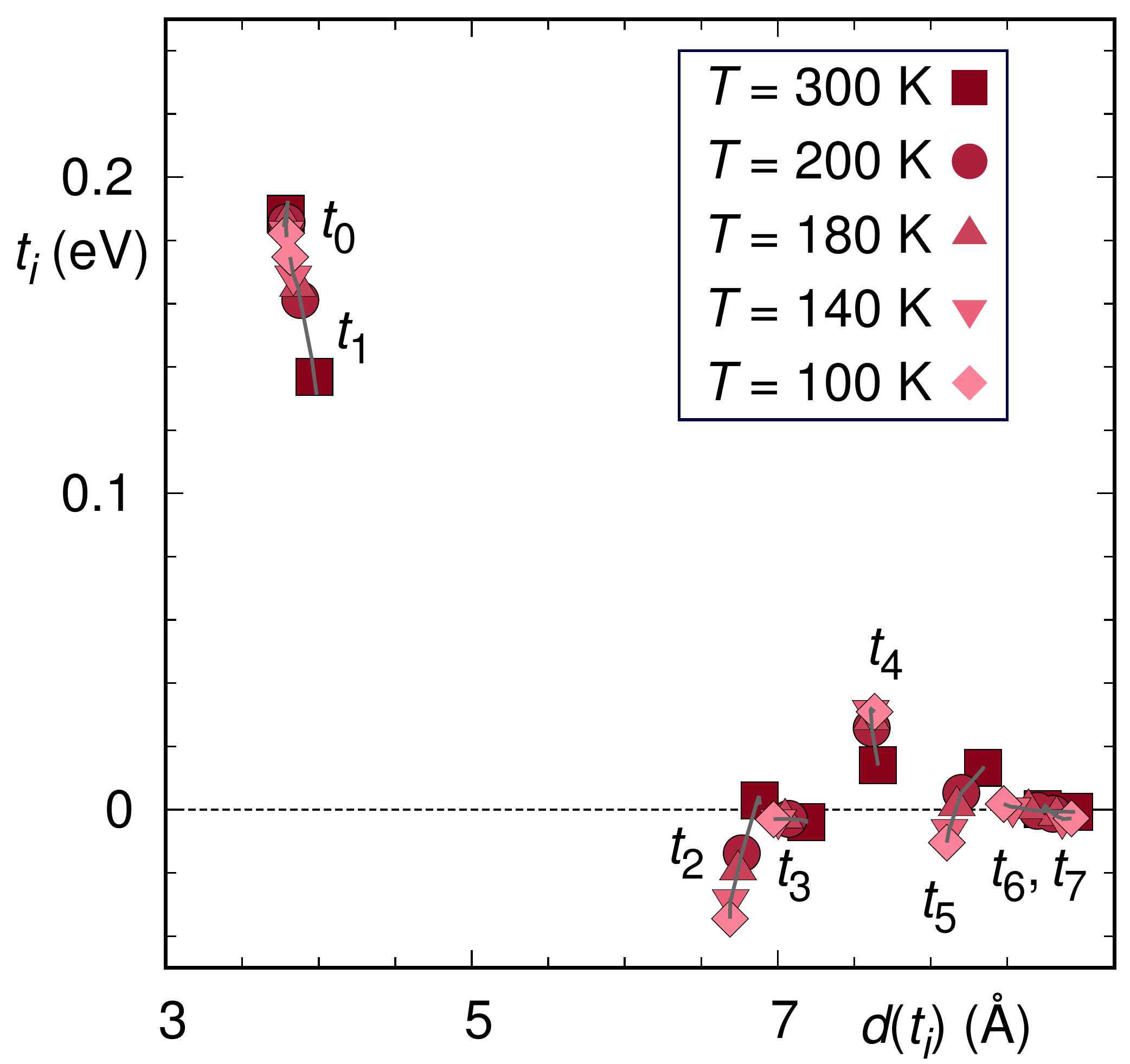}
 \caption{Evolution of tight binding parameters of {\sbf} with
   temperature. As temperature is lowered to $T=100$~K, the dominant
   hoppings $t_0$ and $t_1$ become nearly equal, making the TMTTF
   chain nearly isotropic. The sizable 2D couplings $t_2$, $t_4$ and
   $t_5$ show a complicated temperature dependence, with $t_2$ and
   $t_5$ changing sign and $t_4$ increasing considerably as
   temperature is lowered. }\label{TB_temp}
\end{center}
\end{figure}

In Fig.~\ref{fig:bssbf6temp} we present the band structure of {\sbf}
as a function of temperature. We observe that as the temperature is
decreased, the band width increases and the dispersion between $X$ and
$V$ becomes steeper; this indicates that the electronic structure
becomes more two dimensional with decreasing temperature. This trend
can be also observed in the behavior of the 2D tight-binding parameters
(Fig.~\ref{TB_temp}), especially $t_2$, $t_4$ and $t_5$.

\begin{figure}
\begin{center}
 \includegraphics[width=0.95\columnwidth]{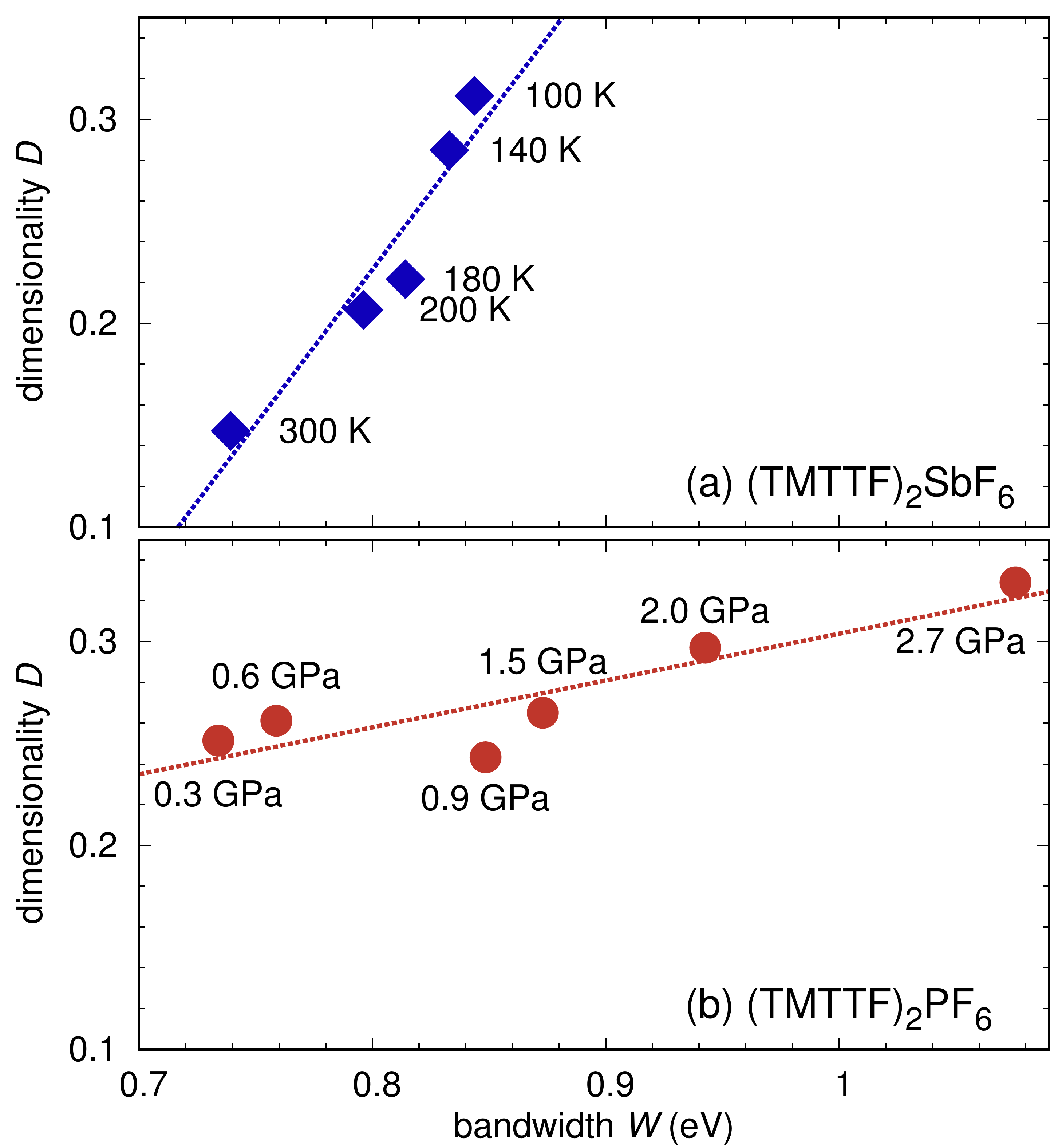} 
 \caption{Dimensionality versus bandwidth of {\sbf} with temperature
   (diamonds) and {\pf} under pressure (circles); the
   bandwidth is the energy difference between the highest and lowest
   energies in the TMTTF bands around the Fermi energy, and the
   dimensionality is defined by the ratio of the hopping integrals in
   the intra- and inter-chain directions (see Eq.~\eqref{eq:Dimension}). We
   see the expected positive correlation between bandwidth and
   pressure (indicated by the arrow): as the pressure increases, so
   does the intermolecular hopping and therefore the bandwidth. We
   also see a strong positive correlation between pressure and
   dimensionality. There is a negative correlation between temperature
   and dimensionality and bandwidth; increasing the temperature has a
   similar effect to decreasing the pressure. The 140 K and 300 K
   points for {\sbf} are averaged over the two structures available
   at those temperatures. }\label{fig:dimensionbandwidth}
\end{center}
\end{figure}

In order to quantify the electronic dimensionality, we introduce a
dimensionality parameter, $D$,  by taking the ratio of
the inter-chain hopping terms  ($t_{\alpha}^\perp$) and
 intra-chain hopping terms
($t_{\beta}^\|$),
\begin{equation}\label{eq:Dimension}
 D = \frac{\sum_{\alpha} |{t}_\alpha^{\perp}|}{ \sum_{\beta}|{t}_\beta^{\|}|}.
\end{equation}
We emphasize that this parameter is an estimate of a model
dimensionality. The correlation between temperature, dimensionality,
and bandwidth is seen more clearly by using this parameter, as
illustrated in Fig.~\ref{fig:dimensionbandwidth} (a), where $D$ for
{\sbf} increases with decreasing temperature.

\begin{figure}
\begin{center}
 \includegraphics[width=0.95\columnwidth]{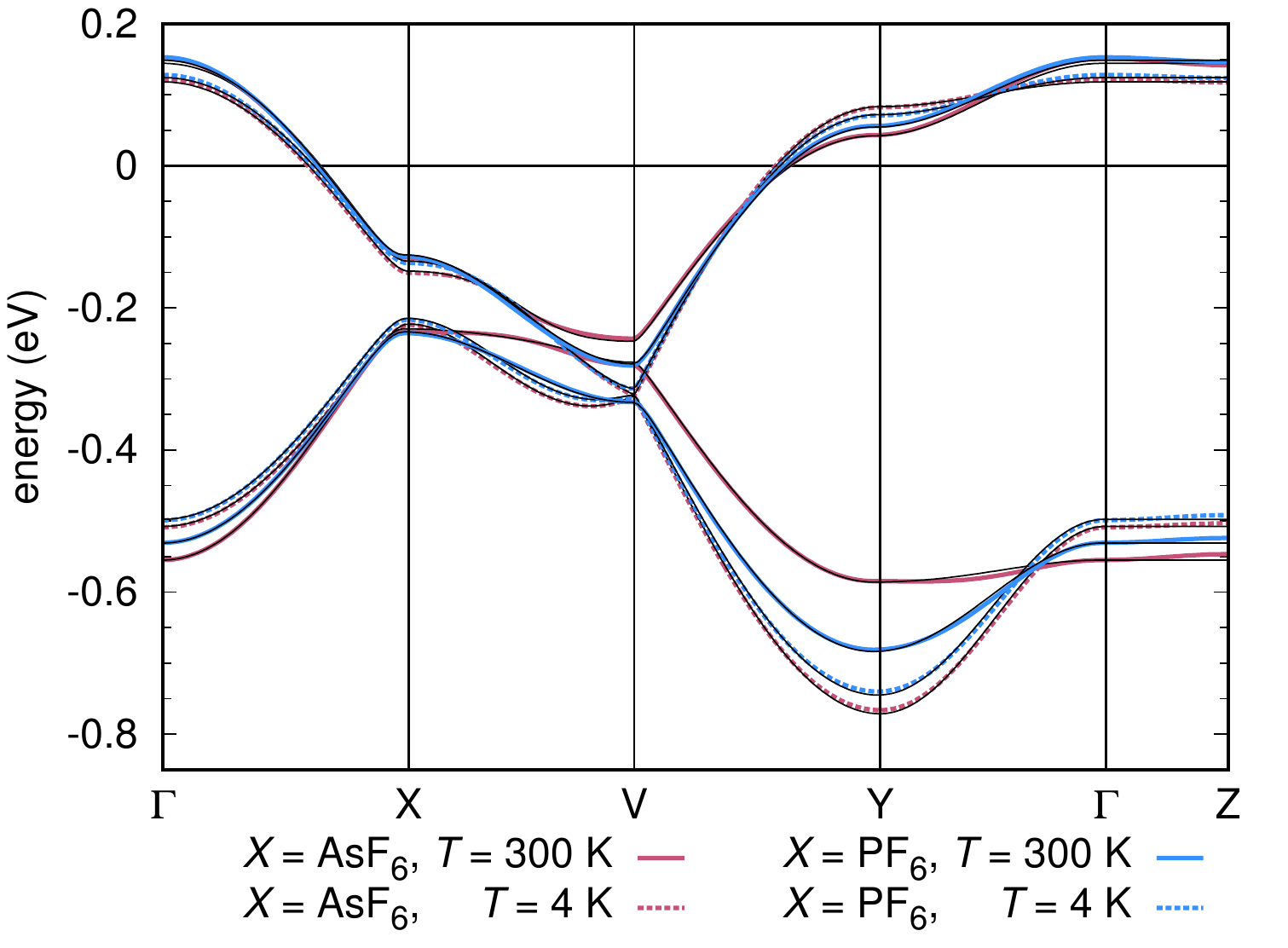}
 \caption{Band structures of {\asf} (red) at $T=4$~K (dashed) and room
   temperature (solid line), {\pf} (blue) at $T=4$~K (dashed) and room
   temperature (solid line). The black lines are the bands resulting
   from the model Hamiltonian Eq.~\eqref{eq:modelH}, parametrized by the
   Wannier orbital overlaps for the 8 shortest hops. In the model used
   there is no dispersion in the $k_z$ direction since the 8 shortest
   hops are all in the same plane, \textit{i.e.} we have a two
   dimensional model. It is clear from the DFT bands that the
   interplanar coupling is small, which is why the 2D model fits so
   well. Note that $T=4$~K is below both the charge ordering and spin
   Peierls transitions of {\asf} and {\pf}.  }\label{fig:asf6pf6}
\end{center}
\end{figure}

In Fig.~\ref{fig:asf6pf6} we present the band structure for {\asf} and
{\pf} for the crystal structures at $T=4$~K and at $T=300$~K. The
inter-chain ($X$-$V$ path) dispersion increases with decreasing 
temperature.  {\asf} and {\pf} undergo spin-Peierls transitions (at
$T=11.4$~K and 16.4~K, respectively),\cite{desouza09} however there is
no energy splitting at $T=4$~K since the crystal structure is not
tetramerized.  Interestingly, the electronic and structural
dimerizations in these systems (see Table~\ref{tab:properties}) are
larger for the room temperature structures than for the structures
measured at $T=4$~K.

We also investigated {\bff}. For this system, the electronic
dimerization changes sign since the electronic dimers are on the more
closely spaced TMTTFs (in terms of center of mass separation) in the room
temperature structure, and on the more distant pair for the structure at 100 K.

\subsection{Pressure dependence of  structural and electronic properties}\label{secVc}

\begin{figure}
\begin{center}
\includegraphics[width=0.95\columnwidth]{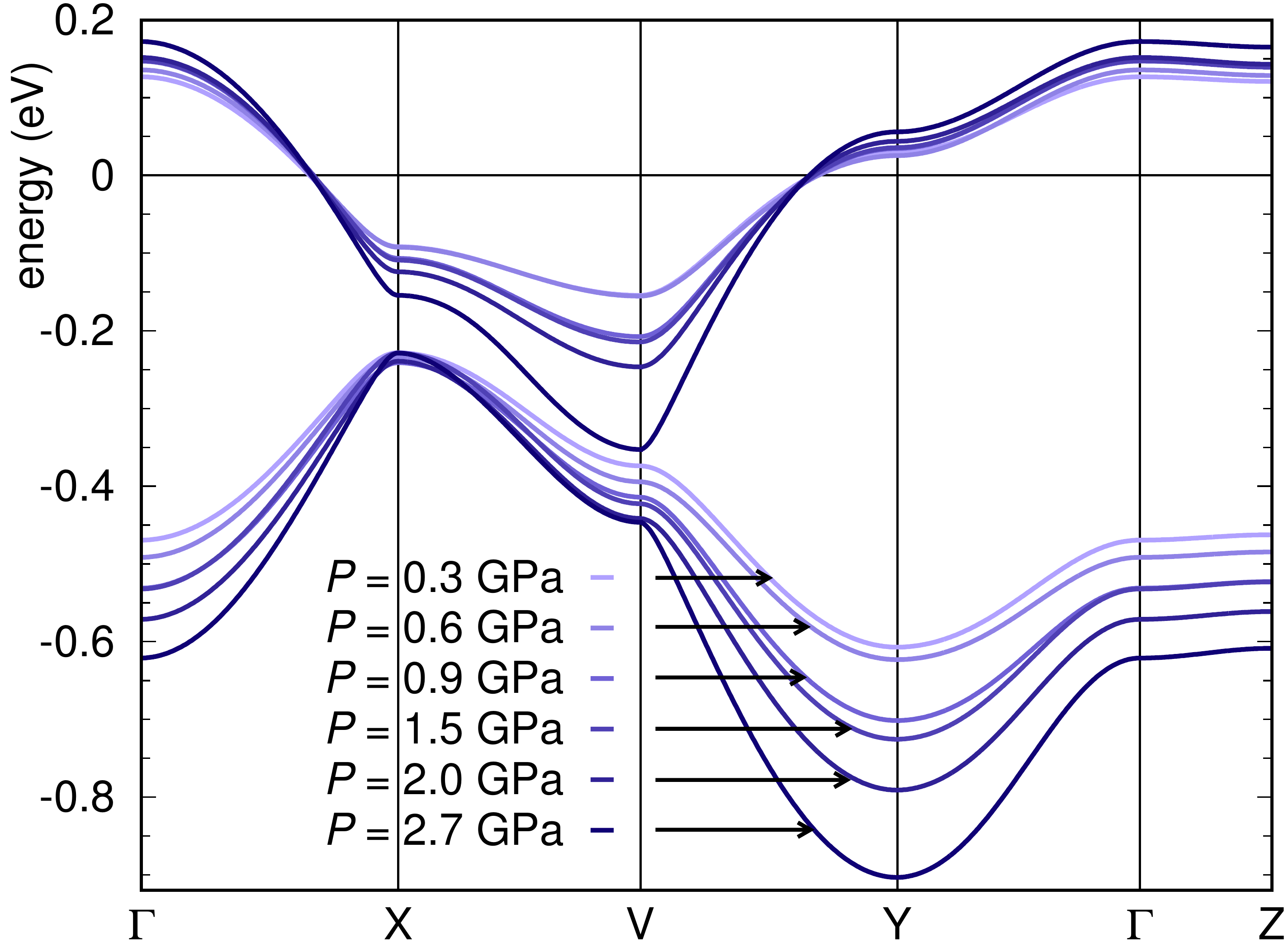} 
\caption{Band structure of {\pf} at various pressures. As the pressure
  is increased, the bandwidth increases, and the dispersion becomes
  steeper between the $X$ and $V$ points; the system becomes more two
  dimensional. These trends are made obvious in
  Figure~\ref{fig:dimensionbandwidth} (b). }\label{fig:bspressure}
\end{center}
\end{figure}

\begin{figure}
\begin{center}
\includegraphics[width=0.95\columnwidth]{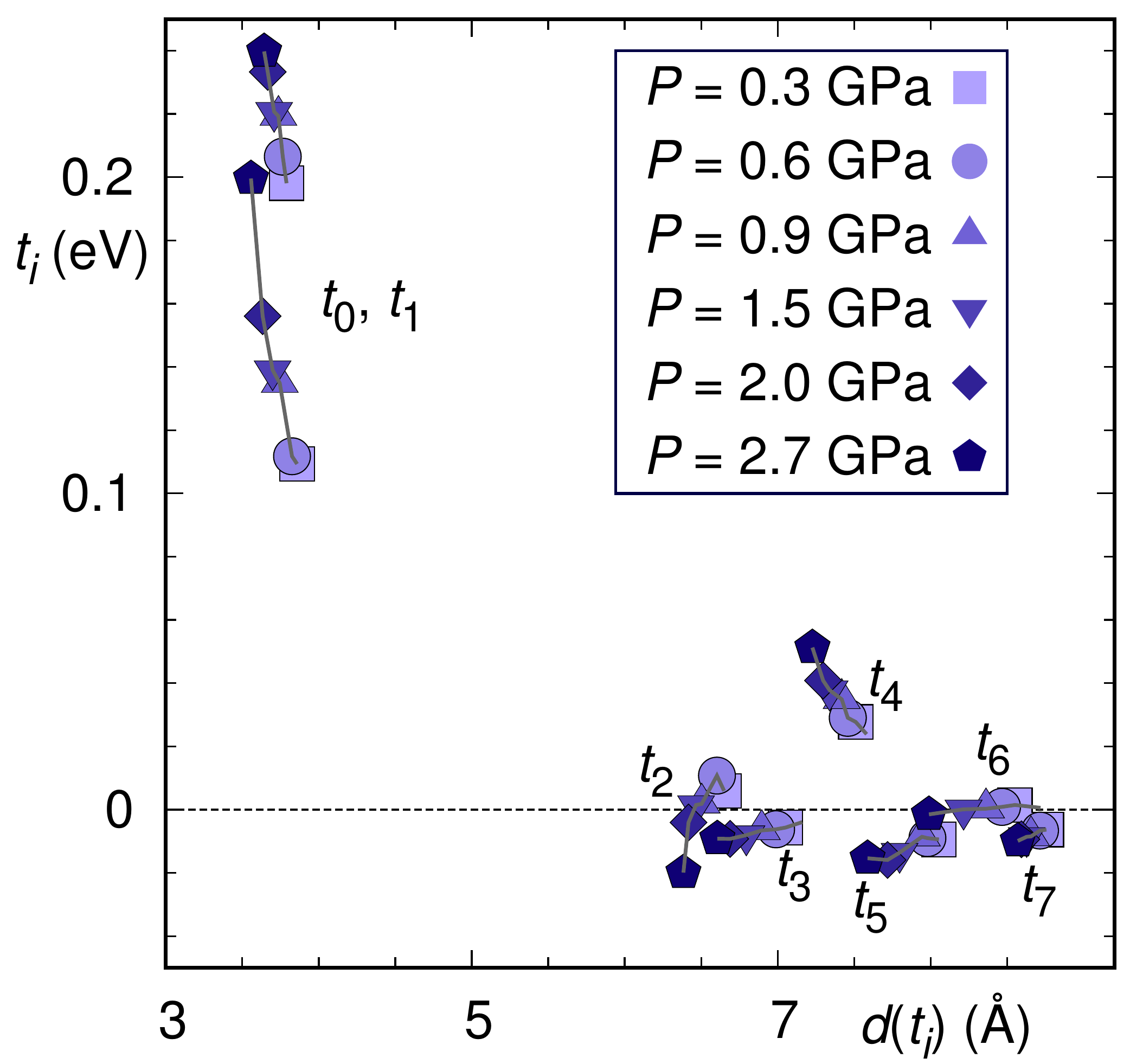} 
\caption{Tight binding parameters of {\pf} at various pressures as a
  function of distance. Note that as the pressure is increased, the
  trend is for the $t$'s to become larger. In
  Figure~\ref{fig:dimensionbandwidth} (b) we see that the increases are
  such that the system becomes more
  two-dimensional.}\label{fig:tpf6pressure}
\end{center}
\end{figure}

Here we investigate a series of new experimental crystal structures of {\pf}
determined at room temperature under various pressures.
Figure~\ref{fig:bspressure} shows how the band structure evolves as a
function of pressure.  As the pressure is increased, the bandwidth
increases, and the system becomes more two-dimensional (\textit{i.e.}
the dispersion is enhanced along the path $X-V$). 
This is also apparent in the tight-binding parameters
(Figure~\ref{fig:tpf6pressure}); all the parameters grow with pressure
(increasing the bandwidth), but not all by the same proportion,
changing the degree of two-dimensionality. This trend to higher
dimensionality has also been observed experimentally - optical
experiments on {\pf} under pressure show that the metallic
conductivity (Drude spectral weight) changes very anisotropically; it
increases quickly with pressure in the perpendicular direction, while
in the in-chain direction there is very little change.\cite{rose13}
This trend with pressure has also been seen in other similar systems
({\asf} and {\pf}), and identified as a cross-over from a
quasi-1D system to a 2D metal.\cite{pashkin06,pashkin10}

Table~\ref{tab:properties} shows that between $P=0.9$~GPa and 1.5~GPa,
the TMTTF molecules become almost equally spaced (in terms of the
centers of mass) since $\delta_{struct} \sim 0$.  At P=1.5~GPa,
 the larger inter-chain hopping is no
longer associated to $t_0$, but to $t_1$; to avoid a discontinuity, we
make an exception to the numbering of $t_\alpha$ with ascending distance
and refer, for pressures 1.5 to 2.7~GPa, to the largest hopping
as $t_0$ even though it belongs to the second nearest neighbour
distance. The inter-chain hoppings $t_0$ and $t_1$ do not become equal
around 1.2~GPa because even when the centers of mass are equally
spaced, the staggering of the molecules in the chain means that the
two hopping integrals are not equivalent.  At pressures above
$P=1.3$~GPa, {\pf} is known experimentally to become metallic, and at low
temperatures undergoes a spin-density wave
transition.\cite{creuzet82,dumm00}

Figure~\ref{fig:dimensionbandwidth} (b) shows how the dimensionality and
bandwidth of {\pf} varies with pressure. We observe the expected
trend of increasing bandwidth under pressure (forcing the TMTTF
molecules closer together, increasing their intermolecular
interactions). We also see that physical pressure changes the
bandwidth more, for a given change in dimensionality.

\section{Model calculations}\label{secVI}

\subsection{Exact Diagonalization of an Extended Hubbard Model}\label{secVIa}

In the previous section we obtained the network of interactions
relevant for the Fabre CT salts by means of DFT calculations. We
proceed now with model calculations in order to analyze the
effect of correlations in these materials.

Since some of the phases realized in these materials are charge 
and spin ordered phases, we shall investigate  charge and spin
structure factors
 using a quarter (hole) filled extended Hubbard model,
\begin{eqnarray}
\label{eq:EHM}
H&=&-\sum_{\langle ij \rangle_8,\sigma} t_{ij} (c^\dag_{i\sigma}c_{j\sigma}+c^\dag_{j\sigma}c_{i\sigma})\nonumber\\
&&+U\sum_i  n_{i\uparrow}n_{i\downarrow} + \sum_{\langle ij \rangle_8} V_{ij} n_{i} n_{j},
\end{eqnarray}
where the sum over $\langle ij \rangle_8$ is over the 8 shortest 
distances between sites,
$ t_{ij}$ (see Eq.~\ref{t_hop}) are the corresponding hopping integrals, $c^\dag_{i\sigma}$
($c_{i\sigma}$) is the creation (annihilation) operator of a hole on
the $i^{th}$ site with spin $\sigma$, and
$n_i=n_{i\uparrow}+n_{i\downarrow}$ with
$n_{i\sigma}=c^\dag_{i\sigma}c_{i\sigma}$. $U$ and $V_{ij}$ are the
on-site and the inter-site Coulomb interactions.

\begin{figure}
\begin{center}
 \includegraphics[width=\columnwidth]{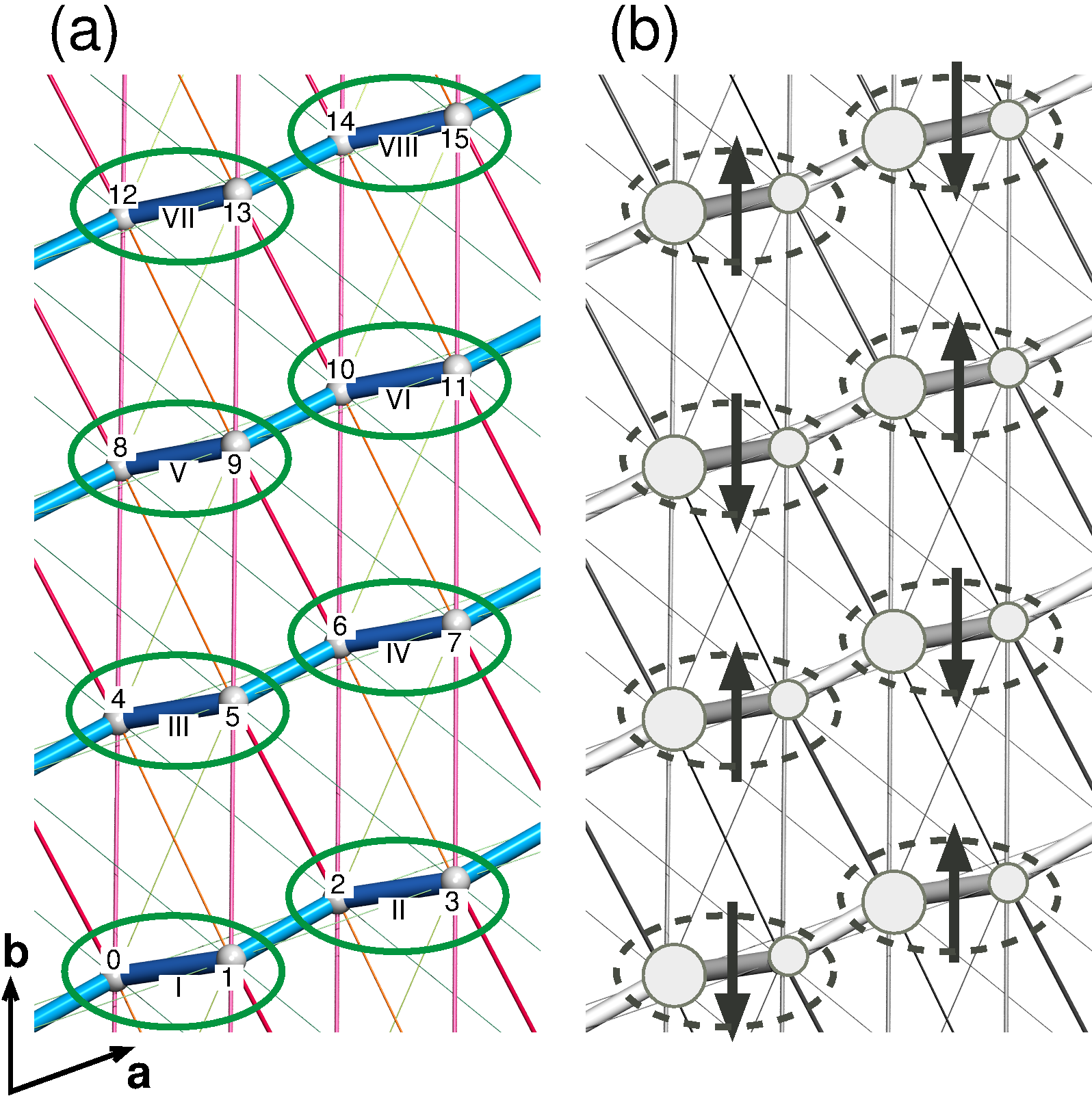}  
 \caption{Schematic representation of the TMTTF molecules in the
   conducting plane for the Fabre salts here studied. (a) Balls
   correspond to the TMTTF molecules and ellipses indicate the
   dimers. The site and dimer labeling is shown.  (b) A sketch of
   a possible charge- and magnetically-ordered state; the size of the
   circles represent the hole density, and the arrows the net
   magnetism. }\label{fig:sheme-TMTTF16}
\end{center}
\end{figure}

In this Section we present numerical results for the Hamiltonian
Eq.~\eqref{eq:EHM}, using the electron hopping parameters ($t_{\alpha}$
 with
$\alpha =0,...,7$) as defined in Eq.~\ref{t_hop} and 
obtained in Sec. V.  We choose the on-site Coulomb
interaction to be typical for this class of materials $U= 4 t_0\approx
1$~eV ($U$ is of the order of the bandwidth $W \sim
1$~eV).\cite{seo06} Since the arrangement of the molecules changes only
slightly with pressure and temperature, we assume that the primary
changes to the intersite interaction are based on the distance between
the sites.  This allows us to reduce the number of free parameters in
our model; we scale $V_{\alpha}$ as a function of the distance $r_{\alpha}$,
$V_{\alpha}=V_0 \frac{r_0}{r_{\alpha}}$, where the index $\alpha$
 corresponds to the label of the
hopping parameter between that pair of sites.  The intersite
interactions $V_7$ and $V_3$ are set to zero since we expect these terms to be strongly
screened by the intermediate sites. Thus we only have two `free'
parameters remaining, $U$ and $V_0$.   The Coulomb interaction along
the chain $V_0$ has been estimated in a previous work as between $0.2 U$ and
$0.6 U$.\cite{seo06}  Here we consider two cases, $ V_0=0.5t_0 $
(weak intersite Coulomb repulsion) and $
V_0=2t_0 $ (strong intersite Coulomb repulsion), both using $U= 4t_0$.
With this set of parameters, the ground state for a system of size $N=16$
(4$\times$4) sites with periodic boundary conditions (see
Figure~\ref{fig:sheme-TMTTF16}) is found using exact diagonalization, as
implemented in ALPS.\cite{alps_bauer11,alps_albuquerque07} 
While similar methods have been applied to some members of this family of materials before,
we note that calculating our parameters from Wannier orbitals allows us to have a more complete, 
realistic description of the inter-chain coupling.\cite{yoshimi12b,yoshimi12}
 
We compute dimer structure factors for charge and spin,
\begin{eqnarray}
  C_D({\bf q})&=&\frac{1}{N_d} \sum_{I,J}\langle n_I n_J \rangle e^{i {\bf q}\cdot({\bf r}_J-{\bf r}_I)}\\
  &&\mbox{with }  n_I=(n_i- n_{i+1})/2\nonumber,
  \end{eqnarray} 
  \begin{eqnarray}
  M_D({\bf q})&=&\frac{1}{N_d} \sum_{I,J}\langle m_I m_J \rangle e^{i {\bf q}\cdot({\bf r}_J-{\bf r}_I)}\\
    &&\mbox{with }  m_I=(m_i+ m_{i+1})/2\nonumber,
\end{eqnarray} 
where $I,J$ are the dimer indices with $i=2 (I-1)$ (and $i$ and
$i+1$ are the site (monomer) indices, see
Figure~\ref{fig:sheme-TMTTF16}), $N_d$ is the total number of dimers,
${\bf r}_I$ denotes the position of the $I^{th}$ dimer, $n_I$ is the
charge difference between the sites in the dimer, and $m_I$ is the
total magnetization of dimers with $m_i=n_{i\uparrow}-n_{i\downarrow}$
the local magnetization.\cite{yoshimi12b,yoshimi12} Note that $C_D$ quantifies the correlation between the charge polarization of dimers, while $M_D$ measures the correlation between  spins on  dimers.

For the $U$ and $V$ values considered here, $C_D({\bf q})$ has a maximum at
${\bf q}=(0,0)$ that corresponds to a charge
order as shown in Fig.~\ref{fig:sheme-TMTTF16} (b).  $M_D({\bf q})$ has a maximum
at ${\bf q}=(\pi,\pi)$ corresponding to a dimer antiferromagnetic
order in both the in-chain and inter-chain directions (shown schematically in Fig.~\ref{fig:sheme-TMTTF16}
(b)). \cite{yoshimi12b,yoshimi12}

\begin{figure}
\begin{center}
 \includegraphics[width=1\columnwidth]{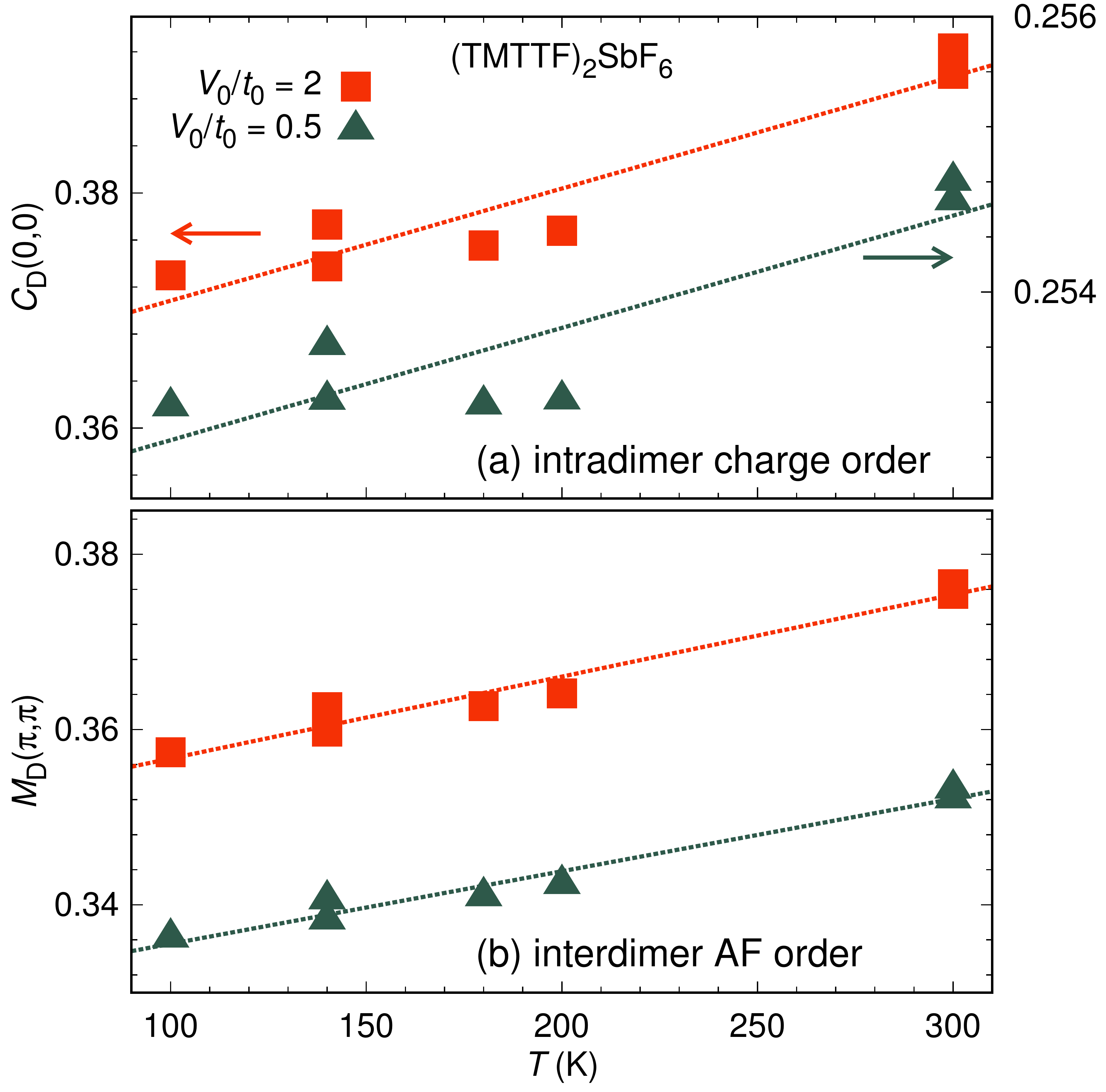}  
 \caption{Structure factors as a function of temperature for
   {\sbf}. As the temperature is decreased the dimer charge and
   magnetic orders are somewhat suppressed. Simultaneously, the
   bandwidth and dimensionality increase. Thus decreasing temperature
   has the same effects as increasing pressure (see
   Figure~\ref{fig:Cmoins-Splus_P}); the charge order is strongly
   activated by $V_0$, and the magnetic order is only weakly
   enhanced. The squares correspond to results with $V_0=2t_0$ while the triangles
   correspond to $V_0 = 0.5 t_0$.}\label{fig:Cmoins-Splus_T}
\end{center}
\end{figure}

\begin{figure}
\begin{center}
 \includegraphics[width=1\columnwidth]{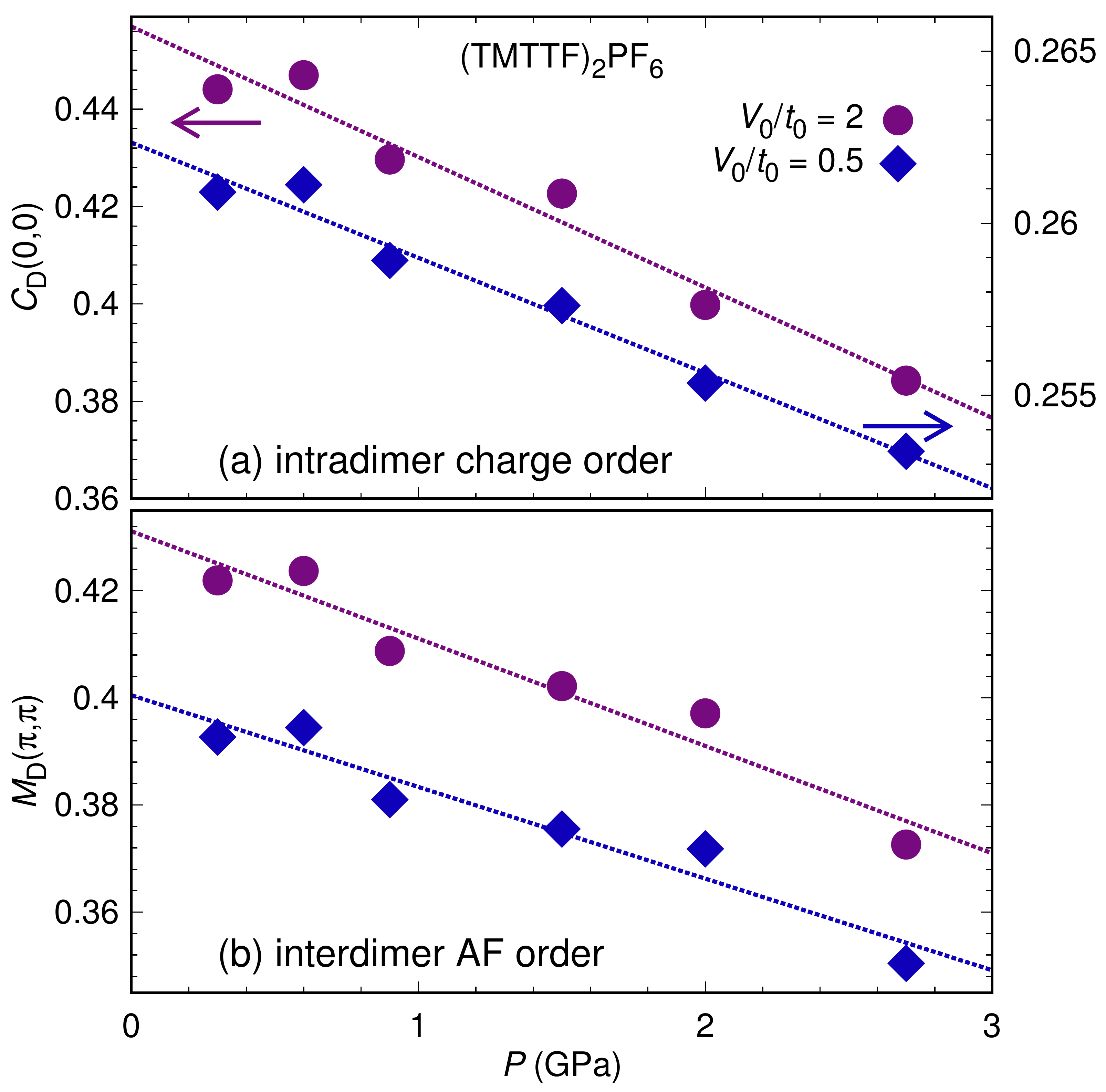}  
 \caption{Structure factors as a function of pressure for {\pf}. As
   the pressure is increased the dimer charge and magnetic orders are
   somewhat suppressed. Simultaneously, the bandwidth and
   dimensionality increase (seen in
   Figure~\ref{fig:dimensionbandwidth} (b)). The charge order is strongly
   activated by the value of $V_0$, while the magnetic order is only weakly
   enhanced, and is even finite for $V_0 = 0$ (not shown). The
    circles correspond to results with $V_0=2t_0$ while the
   diamonds correspond to $V_0 = 0.5 t_0$.}\label{fig:Cmoins-Splus_P}
\end{center}
\end{figure}

\begin{figure}
\includegraphics[width=1\columnwidth]{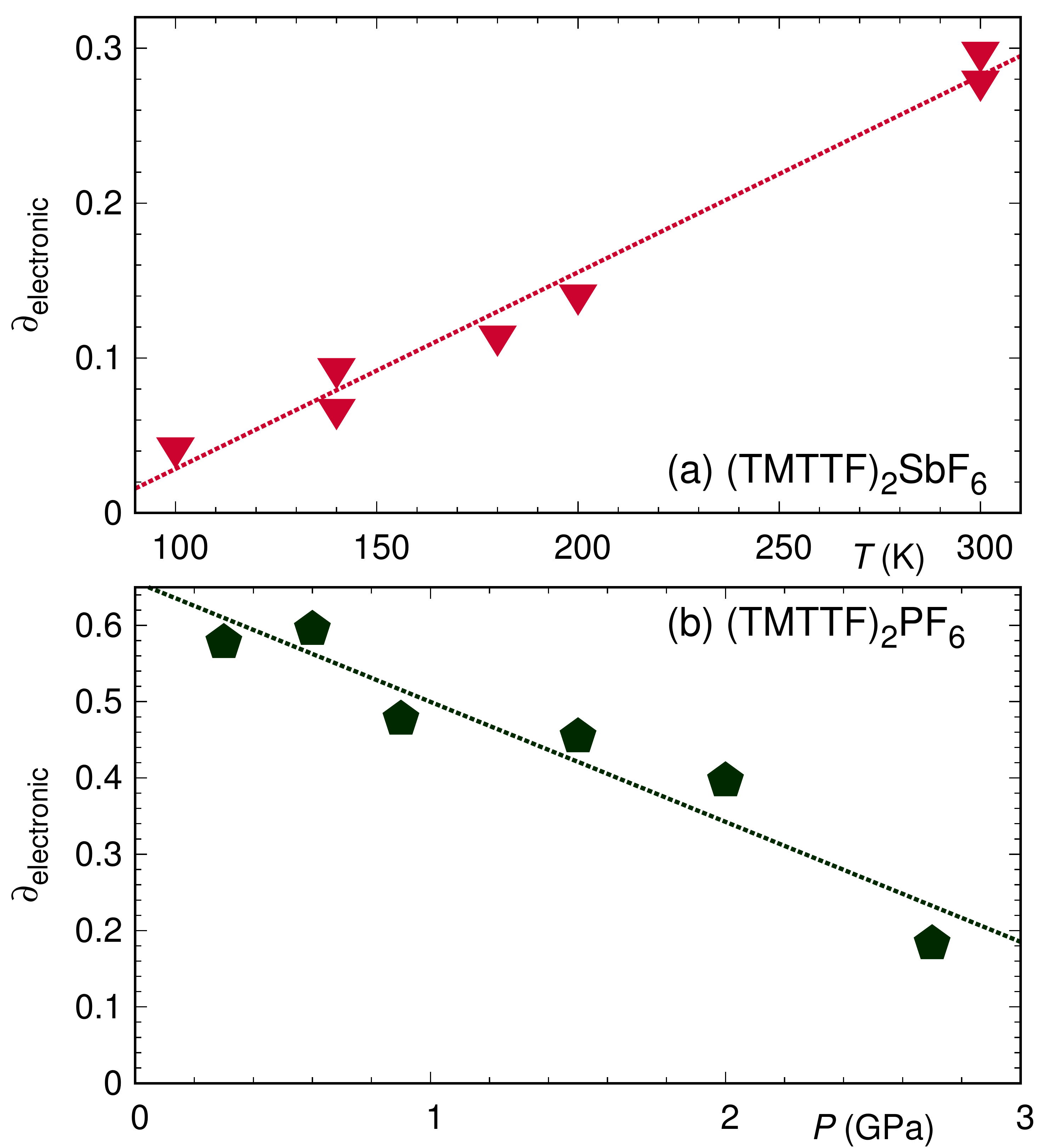}  
 \caption{
Electronic dimerization for (a) {\sbf} as function of temperature and (b) for {\pf} under pressure.
}\label{fig:eldim}
\end{figure}

Here we investigate  charge and spin structure factors for
various Fabre CT structures at different temperatures and pressures.
 To minimize the effects of
experimental variability, we focus our analysis on the sets of
structures synthesized and measured following the same procedure;
 the series of {\pf} under pressure, and {\sbf} for various temperatures.

Both $(0,0)$ charge order and $(\pi,\pi)$
 spin order are slightly suppressed with decreasing temperature
and increasing pressure (Figs.~\ref{fig:Cmoins-Splus_T} 
and \ref{fig:Cmoins-Splus_P}). 
 This is in contrast to a previous work on a simpler
model which showed different trends for the charge and magnetic
orders.\cite{yoshimi12b} Charge order is strongly activated by
increasing the strength of the inter-site Coulomb interaction, $V_0$.  
The antiferromagnetic correlation is relatively weakly enhanced by
increasing $V_0$. 
The changes in the correlation functions shown in
Figs.~\ref{fig:Cmoins-Splus_T} and \ref{fig:Cmoins-Splus_P} seem to be 
strongly connected to the degree of electronic dimerization 
$\partial_{electronic}$ which we show in Fig.~\ref{fig:eldim}. 
Lowering the temperature down to $T=100$~K in {\sbf} continuously 
decreases the electronic dimerization in these structures 
and thus suppresses intra-dimer charge order as well as inter-dimer
antiferromagnetic order by making the one-dimensional chains more
isotropic. The same observation holds for the increase of pressure
on {\pf} structures.

The charge- and magnetically-ordered states  found for the Fabre CT salts
with this model are consistent with the phase diagram for these materials.
But we find no evidence of a phase transition as a function of pressure or
intersite $V$'s in this model. Note that this model cannot capture a
spin-Peierls transition since no magnetoelastic coupling has been 
considered in the Hamiltonian.

In this analysis we have concentrated on the $(0,0)$ charge order and $(\pi,\pi)$
spin order. However, in principle, there may be many kinds of charge order in these
systems. For example, a maximum at $(0,\pi)$ would indicate a charge
order that alternates in the $b$ direction as well as in the $a$
direction. Within the realistic parameter range explored here, we only
observe the type illustrated in Fig.~\ref{fig:sheme-TMTTF16} (b).

\section{Discussion}\label{secVII}

Summarizing our results, we observe that 
the dominant band structure parameters obtainedin our work are generally
consistent with those published for similar materials, however
Ref. \onlinecite{nogami05} finds quite different values for the
electronic dimerization. Those authors compute the $t$ values by
constructing TMTTF HOMOs from an extended H\"uckel model (a
tight-binding model for both the $\sigma$- and $\pi$-bonding systems
of a molecule) and calculate the overlaps between them. This method
does not allow for the charge reorganization and other effects in the
crystal, which are better included by using the Wannier orbitals from
DFT.

Table~\ref{tab:properties} shows that while structural dimerization
tends to decrease with increasing pressure (both chemical and
physical), the electronic dimerization only shows such a trend with
physical pressure; there is no clear trend in electronic dimerization
versus chemical pressure. Under chemical pressure, many aspects of the
molecular arrangement can change (such as spacing and staggering), and
there is no guarantee that they change smoothly with any one parameter
of the anion, such as volume. This is clearest in the series of anions
(SbF$_6$)$^-$, (AsF$_6$)$^-$, (PF$_6$)$^-$ at room temperature. As the
anion size decreases (equivalent to increasing chemical pressure) the
electronic dimerization shows no trend, while the structural
dimerization decreases. It is important to note that the
dimensionality (also computed from the electronic hopping parameters)
shows a clear trend of increasing with increasing chemical and
physical pressure; a trend that has been observed experimentally
in {\pf} and {\asf}.\cite{pashkin06,rose13} 
As the temperature of {\sbf} is increased, it shows a clear increase
in both structural and electronic dimerization.

Our model calculations show that while charge order is strongly
activated by the inter-site Coulomb interaction, $V$, the magnetic
order is weakly enhanced. We also see a weak suppression of both kinds
of order as the pressure is increased, and as the temperature is
decreased.

To reproduce and understand the full phase diagram of these strongly
correlated materials, one needs estimates of the Coulomb
parameters. It is well known that molecular Coulomb parameters are
overestimates for organic crystals; within the crystal, the
interactions are strongly
screened.\cite{scriven09,scriven09B,canocortes10} There are several
promising approaches to calculating the screened Coulomb parameters,
each with their own costs and benefits.\cite{miyake08,canocortes10}
This will be addressed in a future work.

\section{Conclusions}\label{secVIII}

We have  examined the structural and electronic properties
of a set of  Fabre charge transfer salts with crystal structures measured at
different temperatures and pressures. By considering {\it ab initio}
density functional theory calculations we obtain a comparable set of physically meaningful electron
hopping parameters. In these results we identify some general trends: the
structural dimerization is higher for the room temperature systems,
the electronic dimerization decreases with increasing pressure, the
systems are more two dimensional at lower temperatures and higher
pressures, and this change in dimensionality is reflected in the
degree of order in our model Hamiltonian.
With this set of parameters, one can systematically investigate the
differences between these materials in a model Hamiltonian.

It is possible that the variations seen in the electronic structure
(such as the change in dimensionality) are responsible for tuning the
ground states through the various phases accessible in these
materials. However, ideally one would like a similarly systematic set
of many-body interaction parameters, as well as the one-body
parameters given here.

\appendix

\section{Tight binding models}\label{appTB}

In Table~\ref{tab:hoppings} we list all tight binding parameters we
obtained from Wannier function overlaps. 

\begin{table*}
\begin{tabular}{|r|c|cccccccc|c|}
\hline
Anion & $\mu$ & $t_0$ & $t_1$ & $t_2$ & $t_3$ & $t_4$ & $t_5$ & $t_6$ & $t_7$ &optimized\\ \hline
SbF$_6$ 100\,K	&-0.2516 & 0.1823 & 0.1747 & -0.0344 & -0.0030 & 0.0310 & -0.0104 & 0.0018 & -0.0027 &yes\\
SbF$_6$ 140\,K (1) &-0.2459 & 0.1810 & 0.1692 & -0.0285 & -0.0029 & 0.0314 & -0.0061 & 0.0009 & -0.0027 &yes\\
SbF$_6$ 140\,K (2) & -0.2486 & 0.1853 & 0.1687 & -0.0279 & -0.0029 & 0.0318 & -0.0059 & 0.0008 & -0.0030 &yes\\
SbF$_6$ 180\,K 	& -0.2449 & 0.1853 & 0.1652 & -0.0193 & -0.0029 & 0.0283 & 0.0010 & 0.0001 & -0.0020 &yes\\
SbF$_6$ 200\,K 	& -0.2417 & 0.1857 & 0.1612 & -0.0138 & -0.0029 & 0.0259 & 0.0053 & -0.0004 & -0.0013 &yes\\
SbF$_6$ 300\,K (1) & -0.2379 & 0.1947 & 0.1470 & 0.0023 & -0.0032 & 0.0176 & 0.0146 &  $< \pm 10^{-4}$ & -0.0010&yes \\ 
SbF$_6$ 300\,K (2) & -0.2342 & 0.1925 & 0.1426 & 0.0043 & -0.0034 & 0.0140 & 0.0129 & 0.0012 & -0.0007 &yes\\ \hline  
AsF$_6$ 4\,K 	& -0.2617 & 0.1943 & 0.1759 & -0.0380 & -0.0038 & 0.0358 & -0.0141  &  0.0015 & -0.0051 &no\\ 
PF$_6$ 4\,K 	& -0.2539 & 0.1912 & 0.1686 & -0.0333 & -0.0038 & 0.0367 & -0.0104 & 0.0008 & -0.0052  &no\\ 
AsF$_6$ 300\,K 	& -0.2292 & 0.1751 & 0.1568 & -0.0076 & -0.0047 & 0.0188 & 0.0244 & 0.0010 & -0.0021 &yes\\ 
PF$_6$ 300\,K	& -0.2477 & 0.1976 & 0.1569 & -0.0025 & -0.0141 & 0.0030 &   0.0312 &  -0.0003 & -0.0035 &yes\\
PF$_6$ 300\,K, 0.3\,GPa & -0.2203 & 0.1981 & 0.1093 &  0.0059 & -0.0056 & 0.0278 & -0.0094 &  0.0015 & -0.0063 &yes\\
PF$_6$ 300\,K, 0.6\,GPa & -0.2280 & 0.2065 & 0.1118 &  0.0108 & -0.0062 & 0.0291 & -0.0090 & 0.0009 &  -0.0066 &yes\\
PF$_6$ 300\,K, 0.9\,GPa & -0.2513 & 0.2193 & 0.1348 &  0.0019 & -0.0065 & 0.0349 & -0.0085 & 0.0003 &  -0.0085 &yes\\
PF$_6$ 300\,K, 1.5\,GPa& -0.2550 & 0.2207 & 0.1390 &  0.0015 & -0.0080 & 0.0378 & -0.0136 & 0.0001 & -0.0086  &yes\\
PF$_6$ 300\,K, 2.0\,GPa& -0.2757 & 0.2333 & 0.1561 & -0.0041 & -0.0092 & 0.0408 & -0.0159 & $< \pm 10^{-4}$ & -0.0093 &yes\\
PF$_6$ 300\,K, 2.7\,GPa& -0.3081 & 0.2398 & 0.1996  & -0.0199 & -0.0092 & 0.0513 & -0.0154 & -0.0015 & -0.0099 &yes\\ \hline
Br  300\,K & -0.2215 & 0.1719 & 0.1422 & -0.0270 & -0.0048 & 0.0311 & 0.0025 & -0.0007 & -0.0005 &yes\\  
ClO$_4$ 300\,K (A) & -0.2195 & 0.2017 & 0.1067 & -0.0022 & -0.0042 & 0.0313 & -0.0136 & 0.0010 &  -0.0081 &no\\ 
ClO$_4$ 300\,K (B) & -0.2289 & 0.2017 & 0.1067 & -0.0022 & -0.0050 & 0.0320 & -0.0136 & 0.0011 &  -0.0081 &no\\ 
BF$_4$ 100\,K (A) & -0.2435 & 0.1735 & 0.1644 & -0.0151 & -0.0063 & 0.0296 & -0.0240 & 0.0004  &  0.0210 &no\\ 
BF$_4$ 100\,K (B) & -0.2353 & 0.1735 & 0.1644 & -0.0151 & -0.0062 & 0.0300 & -0.0240 & 0.0015 &  0.0210 &no\\ 
BF$_4$ 300\,K (A) & -0.2618 & 0.2057 & 0.1466 & -0.0209 & -0.0010 & 0.0381 & -0.0220 & -0.0013  & -0.0032 &no\\ 
BF$_4$ 300\,K (B) & -0.2413 & 0.2057 & 0.1466 & -0.0209 & -0.0007 & 0.0373 & -0.0220 & -0.0013 & -0.0032  &no\\ \hline
\end{tabular} 
\caption{$t_\alpha$  values determined from the Wannier orbitals for all TMTTF structures investigated here 
  (energies in eV). The $t_\alpha$ are numbered from shortest to longest bond (defined by distance between 
  the centres of mass of the TMTTF molecules), except where noted below. It is clear that the intra-chain terms, $t_0$ and $t_1$, 
  are the dominant hopping terms. 
  Note that for {\pf} above 0.9~GPa, for Br at room temperature, and for BF$_4$ at 100 K the dominant in-chain $t$ is the longer one; 
  we have swapped the labels for these materials such that $t_0$ remains the strongest in-chain coupling.
  The labels (A) and (B) refer to the two inequivalent TMTTFs in each unit cell (in the absence 
  of inversion symmetry).}\label{tab:hoppings}
\end{table*}


\section*{References}

\begin{thebibliography}{44}
\expandafter\ifx\csname natexlab\endcsname\relax\def\natexlab#1{#1}\fi
\expandafter\ifx\csname bibnamefont\endcsname\relax
  \def\bibnamefont#1{#1}\fi
\expandafter\ifx\csname bibfnamefont\endcsname\relax
  \def\bibfnamefont#1{#1}\fi
\expandafter\ifx\csname citenamefont\endcsname\relax
  \def\citenamefont#1{#1}\fi
\expandafter\ifx\csname url\endcsname\relax
  \def\url#1{\texttt{#1}}\fi
\expandafter\ifx\csname urlprefix\endcsname\relax\def\urlprefix{URL }\fi
\providecommand{\bibinfo}[2]{#2}
\providecommand{\eprint}[2][]{\url{#2}}

\bibitem[{\citenamefont{J\'erome}(1991)}]{jerome91}
\bibinfo{author}{\bibfnamefont{D.}~\bibnamefont{J\'erome}},
  \bibinfo{journal}{Science} \textbf{\bibinfo{volume}{252}},
  \bibinfo{pages}{1509} (\bibinfo{year}{1991}).

\bibitem[{\citenamefont{Mori}(2006)}]{mori06}
\bibinfo{author}{\bibfnamefont{H.}~\bibnamefont{Mori}}, \bibinfo{journal}{J.
  Phys. Soc. Jpn.} \textbf{\bibinfo{volume}{75}}, \bibinfo{pages}{051003}
  (\bibinfo{year}{2006}).

\bibitem[{\citenamefont{Yasuzuka and Murata}(2009)}]{yasuzuka09}
\bibinfo{author}{\bibfnamefont{S.}~\bibnamefont{Yasuzuka}} \bibnamefont{and}
  \bibinfo{author}{\bibfnamefont{K.}~\bibnamefont{Murata}},
  \bibinfo{journal}{Sci. Technol. Adv. Mater.} \textbf{\bibinfo{volume}{10}},
  \bibinfo{pages}{024307} (\bibinfo{year}{2009}).

\bibitem[{\citenamefont{Kawakami et~al.}(2003)\citenamefont{Kawakami,
  Taniguchi, Nakano, Kitagawa, and Yamaguchi}}]{kawakami03}
\bibinfo{author}{\bibfnamefont{T.}~\bibnamefont{Kawakami}},
  \bibinfo{author}{\bibfnamefont{T.}~\bibnamefont{Taniguchi}},
  \bibinfo{author}{\bibfnamefont{S.}~\bibnamefont{Nakano}},
  \bibinfo{author}{\bibfnamefont{Y.}~\bibnamefont{Kitagawa}}, \bibnamefont{and}
  \bibinfo{author}{\bibfnamefont{K.}~\bibnamefont{Yamaguchi}},
  \bibinfo{journal}{Polyhedron} \textbf{\bibinfo{volume}{22}},
  \bibinfo{pages}{2051} (\bibinfo{year}{2003}).

\bibitem[{\citenamefont{Yu et~al.}(2004)\citenamefont{Yu, Zhang, Zamborszky,
  Alavi, Baur, Merlic, and Brown}}]{yu04}
\bibinfo{author}{\bibfnamefont{W.}~\bibnamefont{Yu}},
  \bibinfo{author}{\bibfnamefont{F.}~\bibnamefont{Zhang}},
  \bibinfo{author}{\bibfnamefont{F.}~\bibnamefont{Zamborszky}},
  \bibinfo{author}{\bibfnamefont{B.}~\bibnamefont{Alavi}},
  \bibinfo{author}{\bibfnamefont{A.}~\bibnamefont{Baur}},
  \bibinfo{author}{\bibfnamefont{C.~A.} \bibnamefont{Merlic}},
  \bibnamefont{and} \bibinfo{author}{\bibfnamefont{S.~E.} \bibnamefont{Brown}},
  \bibinfo{journal}{Phys. Rev. B} \textbf{\bibinfo{volume}{70}},
  \bibinfo{pages}{121101} (\bibinfo{year}{2004}).

\bibitem[{\citenamefont{Doiron-Leyraud
  et~al.}(2010)\citenamefont{Doiron-Leyraud, Auban-Senzier, de~Cotret,
  Bechgaard, J\'erome, and Taillefer}}]{doironleyraud10}
\bibinfo{author}{\bibfnamefont{N.}~\bibnamefont{Doiron-Leyraud}},
  \bibinfo{author}{\bibfnamefont{P.}~\bibnamefont{Auban-Senzier}},
  \bibinfo{author}{\bibfnamefont{S.~R.} \bibnamefont{de~Cotret}},
  \bibinfo{author}{\bibfnamefont{K.}~\bibnamefont{Bechgaard}},
  \bibinfo{author}{\bibfnamefont{D.}~\bibnamefont{J\'erome}}, \bibnamefont{and}
  \bibinfo{author}{\bibfnamefont{L.}~\bibnamefont{Taillefer}},
  \bibinfo{journal}{Physica B} \textbf{\bibinfo{volume}{405}},
  \bibinfo{pages}{S265} (\bibinfo{year}{2010}).


\bibitem[{\citenamefont{H. C. Kandpal, I. Opahle, Y.-Z. Zhang, 
H. O. Jeschke, R. Valent\'\i}(2009)}]{kandpal09}
\bibinfo{author}{\bibfnamefont{K. C.}~\bibnamefont{Kandpal}},
\bibinfo{author}{\bibfnamefont{I.}~\bibnamefont{Opahle}},
\bibinfo{author}{\bibfnamefont{Y.-Z.}~\bibnamefont{Zhang}},
\bibinfo{author}{\bibfnamefont{H. O.}~\bibnamefont{Jeschke}}, \bibnamefont{and}
  \bibinfo{author}{\bibfnamefont{R.}~\bibnamefont{Valent\'\i}},
  \bibinfo{journal}{Phys. Rev. Lett.} \textbf{\bibinfo{volume}{103}},
  \bibinfo{pages}{067004} (\bibinfo{year}{2009}).

\bibitem[{\citenamefont{}(2012)}]{jeschke12}
\bibinfo{author}{\bibfnamefont{H. O.}~\bibnamefont{Jeschke}},
\bibinfo{author}{\bibfnamefont{M.}~\bibnamefont{de Souza}},
\bibinfo{author}{\bibfnamefont{R.}~\bibnamefont{Valent\'\i}},
\bibinfo{author}{\bibfnamefont{R. S.}~\bibnamefont{Manna}}, 
\bibinfo{author}{\bibfnamefont{M.}~\bibnamefont{Lang}},\bibnamefont{and}
  \bibinfo{author}{\bibfnamefont{J. A.}~\bibnamefont{Schlueter}},
  \bibinfo{journal}{Phys. Rev. B} \textbf{\bibinfo{volume}{85}},
  \bibinfo{pages}{035125} (\bibinfo{year}{2012}).
  
  
  \bibitem [{Note1()}]{Note1}%
  \BibitemOpen
  \bibinfo {note} {Single crystals of {(TMTTF)$_2$PF$_6$} and
  {(TMTTF)$_2$SbF$_6$} were grown electrochemically in single
  or double H-type glass cells at room temperature. A constant voltage of 0.9-1.1~V was applied
  between platinum electrodes of approximately 5x10~mm$^2$ and 15x20~mm$^2$
  respectively, resulting in a starting current through the solution of about
  14$~\mu $A. Needle-shaped single crystals of several millimeters in length
  and less than a millimeter in width are ready to harvest in about 1-4 weeks
  for (TMTTF)$_2$SbF$_6$, and 5-10 weeks for (TMTTF)$_2$PF$_6$. \par 
  Structural investigations on (TMTTF)$_2X$ with $X= $PF$_6$
  and SbF$_6$ under ambient conditions and down to 100~K were performed at
  Universit{\"a}t Stuttgart using a Kappa CCD Bruker AXS diffractometer.
  Diffraction angles $\theta $ of 0.41$^{\circ }$-28.28$^{\circ }$ were
  considered while irradiating with a wavelength of $\lambda $=0.71073~\r A. The
  obtained R-values vary from 0.036 to 0.062. \par The pressure
  dependent x-ray diffraction data for {(TMTTF)$_2$PF$_6$} were collected at the
  ID09A beamline of the European Synchrotron Radiation Facility in Grenoble using a
  diamond anvil cell. A wavelength of 0.413~\r A\ was irradiated with a
  diffraction angle $\theta $ of about 25$^{\circ }$. X-ray diffraction
  patterns were collected on an imaging plate MAR345 detector by rotating the
  crystal from $-30^{\circ }$ to $+30^{\circ }$ with $2^{\circ }$ steps, and
  analyzed using the XDS package.\cite {Kabsch93} The refinement of
  the atomic positions was performed by the SHELX software.\cite {pashkin09,rose13} 
  \par Gaseous helium was used as the pressure transmitting
  medium; the pressure in the cell was determined {\protect \it in situ} by the
  ruby luminescence method.\cite {Mao86} Typical
  R-values are in the range of $0.07-0.08$.}\BibitemShut {Stop}%
  

\bibitem[{\citenamefont{Kabsch}(1993)}]{Kabsch93}
\bibinfo{author}{\bibfnamefont{W.}~\bibnamefont{Kabsch}}, \bibinfo{journal}{J.
  Appl. Cryst.} \textbf{\bibinfo{volume}{26}}, \bibinfo{pages}{795}
  (\bibinfo{year}{1993}).
  
 \bibitem[{\citenamefont{Pashkin et~al.}(2009)\citenamefont{Pashkin, Dressel,
  Ebbinghaus, Hanfland, and Kuntscher}}]{pashkin09}
\bibinfo{author}{\bibfnamefont{A.}~\bibnamefont{Pashkin}},
  \bibinfo{author}{\bibfnamefont{M.}~\bibnamefont{Dressel}},
  \bibinfo{author}{\bibfnamefont{S.~G.} \bibnamefont{Ebbinghaus}},
  \bibinfo{author}{\bibfnamefont{M.}~\bibnamefont{Hanfland}}, \bibnamefont{and}
  \bibinfo{author}{\bibfnamefont{C.~A.} \bibnamefont{Kuntscher}},
  \bibinfo{journal}{Synth. Met.} \textbf{\bibinfo{volume}{159}},
  \bibinfo{pages}{2097 } (\bibinfo{year}{2009}).
  
  \bibitem[{\citenamefont{Rose et~al.}(2013)\citenamefont{Rose, Loose, Kortus,
  Pashkin, Kuntscher, Ebbinghaus, Hanfland, Lissner, Schleid, and
  Dressel}}]{rose13}
\bibinfo{author}{\bibfnamefont{E.}~\bibnamefont{Rose}},
  \bibinfo{author}{\bibfnamefont{C.}~\bibnamefont{Loose}},
  \bibinfo{author}{\bibfnamefont{J.}~\bibnamefont{Kortus}},
  \bibinfo{author}{\bibfnamefont{A.}~\bibnamefont{Pashkin}},
  \bibinfo{author}{\bibfnamefont{C.~A.} \bibnamefont{Kuntscher}},
  \bibinfo{author}{\bibfnamefont{S.~G.} \bibnamefont{Ebbinghaus}},
  \bibinfo{author}{\bibfnamefont{M.}~\bibnamefont{Hanfland}},
  \bibinfo{author}{\bibfnamefont{F.}~\bibnamefont{Lissner}},
  \bibinfo{author}{\bibfnamefont{T.}~\bibnamefont{Schleid}}, \bibnamefont{and}
  \bibinfo{author}{\bibfnamefont{M.}~\bibnamefont{Dressel}},
  \bibinfo{journal}{J. Phys.: Condens. Matter} \textbf{\bibinfo{volume}{25}},
  \bibinfo{pages}{014006} (\bibinfo{year}{2013}).

\bibitem[{\citenamefont{Mao et~al.}(1986)\citenamefont{Mao, Xu, and
  Bell}}]{Mao86}
\bibinfo{author}{\bibfnamefont{H.~K.} \bibnamefont{Mao}},
  \bibinfo{author}{\bibfnamefont{J.}~\bibnamefont{Xu}}, \bibnamefont{and}
  \bibinfo{author}{\bibfnamefont{P.~M.} \bibnamefont{Bell}},
  \bibinfo{journal}{J. Geophys. Res.} \textbf{\bibinfo{volume}{91}},
  \bibinfo{pages}{4673} (\bibinfo{year}{1986}).
  
\bibitem[{\citenamefont{Koepernik and Eschrig}(1999)}]{koepernik99}
\bibinfo{author}{\bibfnamefont{K.}~\bibnamefont{Koepernik}} \bibnamefont{and}
  \bibinfo{author}{\bibfnamefont{H.}~\bibnamefont{Eschrig}},
  \bibinfo{journal}{Phys. Rev. B} \textbf{\bibinfo{volume}{59}},
  \bibinfo{pages}{1743} (\bibinfo{year}{1999}).

\bibitem[{\citenamefont{Perdew et~al.}(1996)\citenamefont{Perdew, Burke, and
  Ernzerhof}}]{perdew96}
\bibinfo{author}{\bibfnamefont{J.~P.} \bibnamefont{Perdew}},
  \bibinfo{author}{\bibfnamefont{K.}~\bibnamefont{Burke}}, \bibnamefont{and}
  \bibinfo{author}{\bibfnamefont{M.}~\bibnamefont{Ernzerhof}},
  \bibinfo{journal}{Phys. Rev. Lett.} \textbf{\bibinfo{volume}{77}},
  \bibinfo{pages}{3865} (\bibinfo{year}{1996}).

\bibitem[{\citenamefont{Kresse and Hafner}(1993)}]{kresse93}
\bibinfo{author}{\bibfnamefont{G.}~\bibnamefont{Kresse}} \bibnamefont{and}
  \bibinfo{author}{\bibfnamefont{J.}~\bibnamefont{Hafner}},
  \bibinfo{journal}{Phys. Rev. B} \textbf{\bibinfo{volume}{47}},
  \bibinfo{pages}{558} (\bibinfo{year}{1993}).

\bibitem[{\citenamefont{Kresse and Furthm\"{u}ller}(1996)}]{kresse96}
\bibinfo{author}{\bibfnamefont{G.}~\bibnamefont{Kresse}} \bibnamefont{and}
  \bibinfo{author}{\bibfnamefont{J.}~\bibnamefont{Furthm\"{u}ller}},
  \bibinfo{journal}{Comput. Mat. Sci.} \textbf{\bibinfo{volume}{6}},
  \bibinfo{pages}{15} (\bibinfo{year}{1996}).

\bibitem[{\citenamefont{Bl\"{o}chl}(1994)}]{blochl94}
\bibinfo{author}{\bibfnamefont{P.~E.} \bibnamefont{Bl\"{o}chl}},
  \bibinfo{journal}{Phys. Rev. B} \textbf{\bibinfo{volume}{50}},
  \bibinfo{pages}{17953} (\bibinfo{year}{1994}).

\bibitem[{\citenamefont{Kresse and Joubert}(1999)}]{kresse99}
\bibinfo{author}{\bibfnamefont{G.}~\bibnamefont{Kresse}} \bibnamefont{and}
  \bibinfo{author}{\bibfnamefont{D.}~\bibnamefont{Joubert}},
  \bibinfo{journal}{Phys. Rev. B} \textbf{\bibinfo{volume}{59}},
  \bibinfo{pages}{1758} (\bibinfo{year}{1999}).

\bibitem[{\citenamefont{Grimme}(2006)}]{grimme06}
\bibinfo{author}{\bibfnamefont{S.}~\bibnamefont{Grimme}}, \bibinfo{journal}{J.
  Comput. Chem.} \textbf{\bibinfo{volume}{27}}, \bibinfo{pages}{1787 }
  (\bibinfo{year}{2006}).

\bibitem[{\citenamefont{Granier et~al.}(1988)\citenamefont{Granier, Gallois,
  Ducasse, Fritsch, and Filhol}}]{granier88}
\bibinfo{author}{\bibfnamefont{T.}~\bibnamefont{Granier}},
  \bibinfo{author}{\bibfnamefont{S.}~\bibnamefont{Gallois}},
  \bibinfo{author}{\bibfnamefont{L.}~\bibnamefont{Ducasse}},
  \bibinfo{author}{\bibfnamefont{A.}~\bibnamefont{Fritsch}}, \bibnamefont{and}
  \bibinfo{author}{\bibfnamefont{A.}~\bibnamefont{Filhol}},
  \bibinfo{journal}{Synth. Met.} \textbf{\bibinfo{volume}{24}},
  \bibinfo{pages}{343} (\bibinfo{year}{1988}).

\bibitem[{\citenamefont{Liautard et~al.}(1982)\citenamefont{Liautard, Peytavin,
  Brun, and Maurin}}]{liautard82b}
\bibinfo{author}{\bibfnamefont{B.}~\bibnamefont{Liautard}},
  \bibinfo{author}{\bibfnamefont{S.}~\bibnamefont{Peytavin}},
  \bibinfo{author}{\bibfnamefont{G.}~\bibnamefont{Brun}}, \bibnamefont{and}
  \bibinfo{author}{\bibfnamefont{M.}~\bibnamefont{Maurin}},
  \bibinfo{journal}{Cryst. Struct. Commun.} \textbf{\bibinfo{volume}{11}},
  \bibinfo{pages}{1841} (\bibinfo{year}{1982}).

\bibitem[{\citenamefont{Galign\'e et~al.}(1978)\citenamefont{Galign\'e,
  Liautard, Peytavin, Brun, Fabre, Torreilles, and Giral}}]{galigne78}
\bibinfo{author}{\bibfnamefont{J.~L.} \bibnamefont{Galign\'e}},
  \bibinfo{author}{\bibfnamefont{B.}~\bibnamefont{Liautard}},
  \bibinfo{author}{\bibfnamefont{S.}~\bibnamefont{Peytavin}},
  \bibinfo{author}{\bibfnamefont{G.}~\bibnamefont{Brun}},
  \bibinfo{author}{\bibfnamefont{J.~M.} \bibnamefont{Fabre}},
  \bibinfo{author}{\bibfnamefont{E.}~\bibnamefont{Torreilles}},
  \bibnamefont{and} \bibinfo{author}{\bibfnamefont{L.}~\bibnamefont{Giral}},
  \bibinfo{journal}{Acta. Cryst. B} \textbf{\bibinfo{volume}{34}},
  \bibinfo{pages}{620} (\bibinfo{year}{1978}).

\bibitem[{\citenamefont{Liautard et~al.}(1984)\citenamefont{Liautard, Peytavin,
  Brun, Chasseau, Fabre, and Giral}}]{liautard84}
\bibinfo{author}{\bibfnamefont{B.}~\bibnamefont{Liautard}},
  \bibinfo{author}{\bibfnamefont{S.}~\bibnamefont{Peytavin}},
  \bibinfo{author}{\bibfnamefont{G.}~\bibnamefont{Brun}},
  \bibinfo{author}{\bibfnamefont{D.}~\bibnamefont{Chasseau}},
  \bibinfo{author}{\bibfnamefont{J.~M.} \bibnamefont{Fabre}}, \bibnamefont{and}
  \bibinfo{author}{\bibfnamefont{L.}~\bibnamefont{Giral}},
  \bibinfo{journal}{Acta. Cryst. C} \textbf{\bibinfo{volume}{40}},
  \bibinfo{pages}{1023} (\bibinfo{year}{1984}).

\bibitem[{\citenamefont{Galign\'e et~al.}(1979)\citenamefont{Galign\'e,
  Liautard, Peytavin, Brun, Maurin, Fabre, Torreilles, and Giral}}]{galigne79a}
\bibinfo{author}{\bibfnamefont{J.~L.} \bibnamefont{Galign\'e}},
  \bibinfo{author}{\bibfnamefont{B.}~\bibnamefont{Liautard}},
  \bibinfo{author}{\bibfnamefont{S.}~\bibnamefont{Peytavin}},
  \bibinfo{author}{\bibfnamefont{G.}~\bibnamefont{Brun}},
  \bibinfo{author}{\bibfnamefont{M.}~\bibnamefont{Maurin}},
  \bibinfo{author}{\bibfnamefont{J.~M.} \bibnamefont{Fabre}},
  \bibinfo{author}{\bibfnamefont{E.}~\bibnamefont{Torreilles}},
  \bibnamefont{and} \bibinfo{author}{\bibfnamefont{L.}~\bibnamefont{Giral}},
  \bibinfo{journal}{Acta. Cryst. B} \textbf{\bibinfo{volume}{35}},
  \bibinfo{pages}{1129} (\bibinfo{year}{1979}).

\bibitem[{\citenamefont{Pedron et~al.}(1994)\citenamefont{Pedron, Bozio,
  Meneghetti, and Pecile}}]{pedron94}
\bibinfo{author}{\bibfnamefont{D.}~\bibnamefont{Pedron}},
  \bibinfo{author}{\bibfnamefont{R.}~\bibnamefont{Bozio}},
  \bibinfo{author}{\bibfnamefont{M.}~\bibnamefont{Meneghetti}},
  \bibnamefont{and} \bibinfo{author}{\bibfnamefont{C.}~\bibnamefont{Pecile}},
  \bibinfo{journal}{Phys. Rev. B} \textbf{\bibinfo{volume}{49}},
  \bibinfo{pages}{10893 } (\bibinfo{year}{1994}).

\bibitem[{\citenamefont{Grant}(1982)}]{grant82}
\bibinfo{author}{\bibfnamefont{P.~M.} \bibnamefont{Grant}},
  \bibinfo{journal}{Phys. Rev. B} \textbf{\bibinfo{volume}{26}},
  \bibinfo{pages}{6888} (\bibinfo{year}{1982}).

\bibitem[{\citenamefont{Whangbo et~al.}(1982)\citenamefont{Whangbo, Jr.,
  Haddon, and Wudl}}]{whangbo82}
\bibinfo{author}{\bibfnamefont{M.-H.} \bibnamefont{Whangbo}},
  \bibinfo{author}{\bibfnamefont{W.~M.~W.} \bibnamefont{Jr.}},
  \bibinfo{author}{\bibfnamefont{R.~C.} \bibnamefont{Haddon}},
  \bibnamefont{and} \bibinfo{author}{\bibfnamefont{R.}~\bibnamefont{Wudl}},
  \bibinfo{journal}{Solid State Commun.} \textbf{\bibinfo{volume}{43}},
  \bibinfo{pages}{637} (\bibinfo{year}{1982}).

\bibitem[{\citenamefont{Ducasse et~al.}(1986)\citenamefont{Ducasse, Abderrabba,
  Hoarau, Pesquer, Gallois, and Gaultier}}]{ducasse86}
\bibinfo{author}{\bibfnamefont{L.}~\bibnamefont{Ducasse}},
  \bibinfo{author}{\bibfnamefont{M.}~\bibnamefont{Abderrabba}},
  \bibinfo{author}{\bibfnamefont{J.}~\bibnamefont{Hoarau}},
  \bibinfo{author}{\bibfnamefont{M.}~\bibnamefont{Pesquer}},
  \bibinfo{author}{\bibfnamefont{B.}~\bibnamefont{Gallois}}, \bibnamefont{and}
  \bibinfo{author}{\bibfnamefont{J.}~\bibnamefont{Gaultier}},
  \bibinfo{journal}{J. Phys. C: Solid State Phys.}
  \textbf{\bibinfo{volume}{19}}, \bibinfo{pages}{3805} (\bibinfo{year}{1986}).

\bibitem[{\citenamefont{Pouget and Ravy}(1996)}]{pouget96}
\bibinfo{author}{\bibfnamefont{J.-P.} \bibnamefont{Pouget}} \bibnamefont{and}
  \bibinfo{author}{\bibfnamefont{S.}~\bibnamefont{Ravy}}, \bibinfo{journal}{J.
  Phys. I France} \textbf{\bibinfo{volume}{6}}, \bibinfo{pages}{1501}
  (\bibinfo{year}{1996}).

\bibitem[{\citenamefont{Dumm et~al.}(2000)\citenamefont{Dumm, Loidl, Fravel,
  Starkey, Montgomery, and Dressel}}]{dumm00}
\bibinfo{author}{\bibfnamefont{M.}~\bibnamefont{Dumm}},
  \bibinfo{author}{\bibfnamefont{A.}~\bibnamefont{Loidl}},
  \bibinfo{author}{\bibfnamefont{B.~W.} \bibnamefont{Fravel}},
  \bibinfo{author}{\bibfnamefont{K.~P.} \bibnamefont{Starkey}},
  \bibinfo{author}{\bibfnamefont{L.~K.} \bibnamefont{Montgomery}},
  \bibnamefont{and} \bibinfo{author}{\bibfnamefont{M.}~\bibnamefont{Dressel}},
  \bibinfo{journal}{Phys. Rev. B} \textbf{\bibinfo{volume}{61}},
  \bibinfo{pages}{511 } (\bibinfo{year}{2000}).

\bibitem[{\citenamefont{Yoshimi
  et~al.}(2012{\natexlab{a}})\citenamefont{Yoshimi, Seo, Ishibashi, and
  Brown}}]{yoshimi12}
\bibinfo{author}{\bibfnamefont{K.}~\bibnamefont{Yoshimi}},
  \bibinfo{author}{\bibfnamefont{H.}~\bibnamefont{Seo}},
  \bibinfo{author}{\bibfnamefont{S.}~\bibnamefont{Ishibashi}},
  \bibnamefont{and} \bibinfo{author}{\bibfnamefont{S.~E.} \bibnamefont{Brown}},
  \bibinfo{journal}{Phys. Rev. Lett.} \textbf{\bibinfo{volume}{108}},
  \bibinfo{pages}{096402} (\bibinfo{year}{2012}{\natexlab{a}}).

\bibitem[{\citenamefont{de~Souza et~al.}(2009)\citenamefont{de~Souza, Br\"uhl,
  M\"uller, Foury-Leylekian, Moradpour, Pouget, and Lang}}]{desouza09}
\bibinfo{author}{\bibfnamefont{M.}~\bibnamefont{de~Souza}},
  \bibinfo{author}{\bibfnamefont{A.}~\bibnamefont{Br\"uhl}},
  \bibinfo{author}{\bibfnamefont{J.}~\bibnamefont{M\"uller}},
  \bibinfo{author}{\bibfnamefont{P.}~\bibnamefont{Foury-Leylekian}},
  \bibinfo{author}{\bibfnamefont{A.}~\bibnamefont{Moradpour}},
  \bibinfo{author}{\bibfnamefont{J.-P.} \bibnamefont{Pouget}},
  \bibnamefont{and} \bibinfo{author}{\bibfnamefont{M.}~\bibnamefont{Lang}},
  \bibinfo{journal}{Physica B} \textbf{\bibinfo{volume}{404}},
  \bibinfo{pages}{494} (\bibinfo{year}{2009}).

\bibitem[{\citenamefont{Pashkin et~al.}(2006)\citenamefont{Pashkin, Dressel,
  and Kuntscher}}]{pashkin06}
\bibinfo{author}{\bibfnamefont{A.}~\bibnamefont{Pashkin}},
  \bibinfo{author}{\bibfnamefont{M.}~\bibnamefont{Dressel}}, \bibnamefont{and}
  \bibinfo{author}{\bibfnamefont{C.~A.} \bibnamefont{Kuntscher}},
  \bibinfo{journal}{Phys. Rev. B} \textbf{\bibinfo{volume}{74}},
  \bibinfo{pages}{165118} (\bibinfo{year}{2006}).

\bibitem[{\citenamefont{Pashkin et~al.}(2010)\citenamefont{Pashkin, Dressel,
  Hanfland, and Kuntscher}}]{pashkin10}
\bibinfo{author}{\bibfnamefont{A.}~\bibnamefont{Pashkin}},
  \bibinfo{author}{\bibfnamefont{M.}~\bibnamefont{Dressel}},
  \bibinfo{author}{\bibfnamefont{M.}~\bibnamefont{Hanfland}}, \bibnamefont{and}
  \bibinfo{author}{\bibfnamefont{C.~A.} \bibnamefont{Kuntscher}},
  \bibinfo{journal}{Phys. Rev. B} \textbf{\bibinfo{volume}{81}},
  \bibinfo{pages}{125109} (\bibinfo{year}{2010}).

\bibitem[{\citenamefont{Creuzet et~al.}(1982)\citenamefont{Creuzet, Takahashi,
  and J\'erome}}]{creuzet82}
\bibinfo{author}{\bibfnamefont{F.}~\bibnamefont{Creuzet}},
  \bibinfo{author}{\bibfnamefont{T.}~\bibnamefont{Takahashi}},
  \bibnamefont{and} \bibinfo{author}{\bibfnamefont{D.}~\bibnamefont{J\'erome}},
  \bibinfo{journal}{J. Phys. Lett.} \textbf{\bibinfo{volume}{43}},
  \bibinfo{pages}{755 } (\bibinfo{year}{1982}).

\bibitem[{\citenamefont{Seo et~al.}(2006)\citenamefont{Seo, Merino, Yoshioka,
  and Ogata}}]{seo06}
\bibinfo{author}{\bibfnamefont{H.}~\bibnamefont{Seo}},
  \bibinfo{author}{\bibfnamefont{J.}~\bibnamefont{Merino}},
  \bibinfo{author}{\bibfnamefont{H.}~\bibnamefont{Yoshioka}}, \bibnamefont{and}
  \bibinfo{author}{\bibfnamefont{M.}~\bibnamefont{Ogata}}, \bibinfo{journal}{J.
  Phys. Soc. Jpn.} \textbf{\bibinfo{volume}{75}}, \bibinfo{pages}{051009}
  (\bibinfo{year}{2006}).

\bibitem[{\citenamefont{{B. Bauer \textit{et al}. (ALPS
  collaboration)}}(2011)}]{alps_bauer11}
\bibinfo{author}{\bibnamefont{{B. Bauer \textit{et al}. (ALPS
  collaboration)}}}, \bibinfo{journal}{J. Stat. Mech.}
  \textbf{\bibinfo{volume}{P05001}} (\bibinfo{year}{2011}).

\bibitem[{\citenamefont{{A.F. Albuquerque \textit{et al}. (ALPS
  collaboration)}}(2007)}]{alps_albuquerque07}
\bibinfo{author}{\bibnamefont{{A.F. Albuquerque \textit{et al}. (ALPS
  collaboration)}}}, \bibinfo{journal}{J. Mag. Mag. Mater.}
  \textbf{\bibinfo{volume}{310}}, \bibinfo{pages}{1187} (\bibinfo{year}{2007}).

\bibitem[{\citenamefont{Yoshimi
  et~al.}(2012{\natexlab{b}})\citenamefont{Yoshimi, Seo, Ishibashi, and
  Brown}}]{yoshimi12b}
\bibinfo{author}{\bibfnamefont{K.}~\bibnamefont{Yoshimi}},
  \bibinfo{author}{\bibfnamefont{H.}~\bibnamefont{Seo}},
  \bibinfo{author}{\bibfnamefont{S.}~\bibnamefont{Ishibashi}},
  \bibnamefont{and} \bibinfo{author}{\bibfnamefont{S.~E.} \bibnamefont{Brown}},
  \bibinfo{journal}{Physica B} \textbf{\bibinfo{volume}{407}},
  \bibinfo{pages}{1783 } (\bibinfo{year}{2012}{\natexlab{b}}).

\bibitem[{\citenamefont{Meneghetti et~al.}(1984)\citenamefont{Meneghetti,
  Bozio, Zanon, Pecile, Ricotta, and Zanetti}}]{meneghetti84}
\bibinfo{author}{\bibfnamefont{M.}~\bibnamefont{Meneghetti}},
  \bibinfo{author}{\bibfnamefont{R.}~\bibnamefont{Bozio}},
  \bibinfo{author}{\bibfnamefont{I.}~\bibnamefont{Zanon}},
  \bibinfo{author}{\bibfnamefont{C.}~\bibnamefont{Pecile}},
  \bibinfo{author}{\bibfnamefont{C.}~\bibnamefont{Ricotta}}, \bibnamefont{and}
  \bibinfo{author}{\bibfnamefont{M.}~\bibnamefont{Zanetti}},
  \bibinfo{journal}{J. Chem. Phys.} \textbf{\bibinfo{volume}{80}},
  \bibinfo{pages}{6210 } (\bibinfo{year}{1984}).

\bibitem[{\citenamefont{Bozio et~al.}(1982)\citenamefont{Bozio, Meneghetti, and
  Pecile}}]{bozio82}
\bibinfo{author}{\bibfnamefont{R.}~\bibnamefont{Bozio}},
  \bibinfo{author}{\bibfnamefont{M.}~\bibnamefont{Meneghetti}},
  \bibnamefont{and} \bibinfo{author}{\bibfnamefont{C.}~\bibnamefont{Pecile}},
  \bibinfo{journal}{J. Chem. Phys.} \textbf{\bibinfo{volume}{76}},
  \bibinfo{pages}{5785 } (\bibinfo{year}{1982}).

\bibitem[{\citenamefont{Nogami et~al.}(2005)\citenamefont{Nogami, Ito,
  Yamamoto, Irie, Horita, Kambe, Nagao, Oshima, Ikeda, and
  Nakamura}}]{nogami05}
\bibinfo{author}{\bibfnamefont{Y.}~\bibnamefont{Nogami}},
  \bibinfo{author}{\bibfnamefont{T.}~\bibnamefont{Ito}},
  \bibinfo{author}{\bibfnamefont{K.}~\bibnamefont{Yamamoto}},
  \bibinfo{author}{\bibfnamefont{N.}~\bibnamefont{Irie}},
  \bibinfo{author}{\bibfnamefont{S.}~\bibnamefont{Horita}},
  \bibinfo{author}{\bibfnamefont{T.}~\bibnamefont{Kambe}},
  \bibinfo{author}{\bibfnamefont{N.}~\bibnamefont{Nagao}},
  \bibinfo{author}{\bibfnamefont{K.}~\bibnamefont{Oshima}},
  \bibinfo{author}{\bibfnamefont{N.}~\bibnamefont{Ikeda}}, \bibnamefont{and}
  \bibinfo{author}{\bibfnamefont{T.}~\bibnamefont{Nakamura}},
  \bibinfo{journal}{J. Phys. (Paris) IV} \textbf{\bibinfo{volume}{131}},
  \bibinfo{pages}{39} (\bibinfo{year}{2005}).

\bibitem[{\citenamefont{Scriven and Powell}(2009{\natexlab{a}})}]{scriven09}
\bibinfo{author}{\bibfnamefont{E.}~\bibnamefont{Scriven}} \bibnamefont{and}
  \bibinfo{author}{\bibfnamefont{B.~J.} \bibnamefont{Powell}},
  \bibinfo{journal}{J. Chem. Phys.} \textbf{\bibinfo{volume}{130}},
  \bibinfo{pages}{104508} (\bibinfo{year}{2009}{\natexlab{a}}).

\bibitem[{\citenamefont{Scriven and Powell}(2009{\natexlab{b}})}]{scriven09B}
\bibinfo{author}{\bibfnamefont{E.}~\bibnamefont{Scriven}} \bibnamefont{and}
  \bibinfo{author}{\bibfnamefont{B.~J.} \bibnamefont{Powell}},
  \bibinfo{journal}{Phys. Rev. B} \textbf{\bibinfo{volume}{80}},
  \bibinfo{pages}{205107} (\bibinfo{year}{2009}{\natexlab{b}}).

\bibitem[{\citenamefont{Cano-Cortes et~al.}(2010)\citenamefont{Cano-Cortes,
  Dolfen, Merino, and Koch}}]{canocortes10}
\bibinfo{author}{\bibfnamefont{L.}~\bibnamefont{Cano-Cortes}},
  \bibinfo{author}{\bibfnamefont{A.}~\bibnamefont{Dolfen}},
  \bibinfo{author}{\bibfnamefont{J.}~\bibnamefont{Merino}}, \bibnamefont{and}
  \bibinfo{author}{\bibfnamefont{E.}~\bibnamefont{Koch}},
  \bibinfo{journal}{Physica B} \textbf{\bibinfo{volume}{405}},
  \bibinfo{pages}{S185} (\bibinfo{year}{2010}).

\bibitem[{\citenamefont{Miyake and Aryasetiawan}(2008)}]{miyake08}
\bibinfo{author}{\bibfnamefont{T.}~\bibnamefont{Miyake}} \bibnamefont{and}
  \bibinfo{author}{\bibfnamefont{F.}~\bibnamefont{Aryasetiawan}},
  \bibinfo{journal}{Phys. Rev. B} \textbf{\bibinfo{volume}{77}},
  \bibinfo{pages}{085122} (\bibinfo{year}{2008}).


\end{thebibliography}
\end{document}